\documentclass[letterpaper,12pt]{article}
\usepackage{amsmath,amssymb,euscript}

\def\0b{\ }
\def\one{{\mathchoice {\rm 1\mskip-4mu l} {\rm 1\mskip-4mu} {\rm 1\mskip-4.5mu l} {\rm 1\mskip-5mu l}}}

\def\onefourpi{\frac{1}{4\pi}}
\def\onetwopi{\frac{1}{2\pi}}

\def\a{\alpha}

\def\b{\beta}

\def\be{\begin{equation}}
\def\bea{\begin{eqnarray}}
\def\beas{\begin{eqnarray*}}

\def\c{\gamma}

\def\d{\delta}
\def\D{\Delta}

\def\eb{{\bar{e}}}
\def\ee{\end{equation}}
\def\eea{\end{eqnarray}}
\def\eeas{\end{eqnarray*}}

\def\half{\frac{1}{2}}

\def\L{\Lambda}

\def\nn{\nonumber}

\def\o{\omega}
\def\ob{\bar{\omega}}
\def\p{\pi}
\def\pl{\partial}

\def\sl{{\cal L}}

\def\big#1{\mbox{\large $#1$}}

\def\namegroup#1{\begin{eqnarray}\label{#1}\nonumber\end{eqnarray}\vspace{-.5in}}

\renewcommand{\theequation}{\arabic{section}.\arabic{equation}}

\newcounter{saveeqn}
\newcounter{savealpheqn}

\newcommand{\alpheqn}{\setcounter{saveeqn}{\value{equation}}%
 \stepcounter{saveeqn}\setcounter{equation}{0}%
 \renewcommand{\theequation}{\mbox{\arabic{section}.\arabic{saveeqn}\alph{equation}}}
}
\def\group#1{\refstepcounter{equation}\setcounter{saveeqn}{\value{equation}}%
 \label{#1}\setcounter{equation}{0}%
 \renewcommand{\theequation}{\mbox{\arabic{section}.\arabic{saveeqn}\alph{equation}}}
}
\newcommand{\reseteqn}{\setcounter{equation}{\value{saveeqn}}%
 \renewcommand{\theequation}{\arabic{section}.\arabic{equation}}%
}

\newcommand{\aalpheqn}{\setcounter{saveeqn}{\value{equation}}%
 \stepcounter{saveeqn}\setcounter{equation}{0}%
 \renewcommand{\theequation}{\mbox{\Alph{subsection}.\arabic{saveeqn}\alph{equation}}}
}
       \newcommand{\areseteqn}{\setcounter{equation}{\value{saveeqn}}%
 \renewcommand{\theequation}{\Alph{subsection}.\arabic{equation}}%
}

\def\group#1{\refstepcounter{equation}\setcounter{saveeqn}{\value{equation}}%
 \label{#1}\setcounter{equation}{0}%
 \renewcommand{\theequation}{\mbox{\arabic{section}.\arabic{saveeqn}\alph{equation}}}
}       
\renewcommand{\thefootnote}{\alph{footnote}}

\newcommand{\newsection}{\setcounter{equation}{0}\section}

\newcommand{\appendixa}
 {\renewcommand{\theequation}{\Alph{subsection}.\arabic{equation}}%
  \renewcommand{\thesubsection}%
               {Appendix \Alph{subsection}.\setcounter{equation}{0}}%
  \renewcommand{\alpheqn}{\aalpheqn}%
  \renewcommand{\reseteqn}{\areseteqn}
  \newcounter{savesec}
  \stepcounter{savesec}}
\newcommand{\appendices}{\appendix\appendixa}
\def\app#1#2{\renewcommand{\thesubsection}{\Alph{subsection}}%
        \setcounter{equation}{0} 
        \refstepcounter{subsection}%
        \setcounter{subsection}{\value{savesec}}%
        \stepcounter{savesec}\label{#1}%
        \renewcommand{\thesubsection}%
               {Appendix \Alph{subsection}.\setcounter{equation}{0}}%
        \addcontentsline{toc}{subsection}{Appendix \Alph{subsection}.  #2}
        \subsection*{Appendix \Alph{subsection}.   #2}}

\newcounter{myfigctr}

\begin{document}


\begin{titlepage}
\begin{center}

January 30, 2000                \hfill UCB-PTH-01/01   \\
                                \hfill LBNL-47406    \\
                                \hfill hep-th/0101220 \\

\vskip .75in
\def\thefootnote{\fnsymbol{footnote}}
{\large \bf HAMILTONIAN FORMULATION OF OPEN WZW STRINGS}

\vskip 0.3in

S. Giusto \footnote{On leave from Dipartimento di Fisica dell'Universit\`a di Genova, Via Dodecaneso, 33, I-16146, Genoa, Italy. 
E-Mail: SGiusto@lbl.gov, giusto@ge.infn.it} and M. B. Halpern \footnote{E-Mail: halpern@physics.berkeley.edu} 

\vskip .2in

     {\em Department of Physics, University of California \\
          and\\
     Theoretical Physics Group, Lawrence Berkeley National Laboratory\\
     Berkeley, California 94720, USA}

        
\end{center}

\vskip .3in

\vfill

\begin{abstract}
Using a Hamiltonian approach, we construct the classical and quantum
theory of open WZW strings on a strip. (These are the strings which
end on WZW branes.) The development involves non-abelian generalized
Dirichlet images in an essential way. At the classical level, we find
a new non-commutative geometry in which the equal-time coordinate
brackets are non-zero at the world-sheet boundary, and the result is
an intrinsically non-abelian effect which vanishes in the abelian
limit. Using the classical theory as a guide to the quantum theory, we
also find the operator algebra and the analogue of the
Knizhnik-Zamolodchikov equations for the conformal field theory of
open WZW strings.
\end{abstract}

\vfill

\end{titlepage}
\setcounter{footnote}{0}


\newsection{Introduction}

Affine Lie algebra \cite{KM,BH,Rev} has played a central role in
string theory over the decades, first in open string theory, then
closed string theory and now back again to open strings. We refer in
particular to the theory\footnote{Early papers on Dirichlet boundary
conditions in string theory include Refs.~\cite{CF,Si}.} of D-branes
\cite{Hor,Po}, open descendants [\ref{refSa}-\ref{refPSS3}] and WZW branes
[\ref{refKS}-\ref{refFS}]. The
present paper is concerned with open WZW strings, which end on WZW
branes.

In particular, we shall construct the classical and quantum theory of
open WZW strings on a strip, following the principles:
\begin{itemize}
\item At the quantum level, open string WZW dynamics on the Lie
algebra $g$ must follow from the Hamiltonian
\begin{equation}
H_g=L_g(0)=L_g^{ab}\sum_{m\in\mathbb{Z}} :J_a(m)J_b(-m):
\end{equation}
where the current modes $\{J_a(m)\}$ generate the affine Lie algebra
of $g$ and $L_g(0)$ is the zero mode of the affine-Sugawara
construction [\ref{refBH},~\ref{refHa}-\ref{refSe},~\ref{refRev}] on $g$. Thus, open WZW strings
are controlled by a single non-abelian chiral current.

\item The dynamics of open WZW strings on the strip $0\le\xi\le\pi$
must be locally WZW in the bulk
\begin{equation}
0<\xi<\pi
\end{equation}
that is, the same as the ordinary WZW model \cite{No,Wi}. The
classical group elements $g(T,\xi,t)$ of the open WZW string must
satisfy generalized Dirichlet boundary conditions \cite{AS}
\begin{equation}
g^{-1}(T,\xi,t) \pl_+ g(T,\xi,t) = g(T,\xi,t) \pl_- g^{-1}(T,\xi,t)
\quad {\rm at}\,\, \xi=0,\p
\end{equation}
at the boundary of the strip.
\end{itemize}
As we will see, both the classical and quantum theory involve
generalized non-abelian Dirichlet image charges in an essential way.

At the classical level (see Secs.~2-6), we find that the single non-abelian
chiral current
algebra determines both the phase space and the coordinate space
formulation of open WZW theory.  In the phase space formulation, we
obtain the complete bracket algebra of the theory and, in particular, we
find a new equal-time non-commutative geometry in which the coordinate
brackets
\begin{equation}
\{x^i(\xi,t),x^j(\eta,t)\}=\left\{\begin{array}{ll} \ne 0 & \mbox{if
$\xi=\eta=0$ or $\pi$} \\ 0 & \mbox{otherwise} \end{array} \right .
\end{equation}
are non-zero at the boundary. This effect (which is not related in any 
simple sense to
known [\ref{refCDS}-\ref{refSW}] non-commutative effects) is intrinsically non-abelian
and vanishes in the abelian limit.

Using the classical theory as a guide, the quantum theory of open
strings is constructed in Sec.~7. Here we find the operator algebra of
the theory, the differential equations for the vertex operators and
the analogue of the Knizhnik-Zamolodchikov equations \cite{DF,KZ} for
the conformal field theory of open WZW strings. We mention in
particular that the open string vertex operators $g(T)$ can be
factorized into chiral vertex operators $g_\pm(T)$
\begin{equation}
g(T)=g_-(T)g_+(T),\quad \pl_-g_+(T)=\pl_+ g_-(T)=0
\end{equation} 
but, in distinction to ordinary WZW theory, $g_-(T)$ and $g_+(T)$ do
not form independent subspaces and the open WZW string correlators do
not factorize into left and right mover correlators.
 
\vspace{-.1in}
\newsection{Open WZW Strings}

\vspace{-.07in}
\subsection{Quantum Formulation}

The Hamiltonian $H$ of any open string theory is the zero mode $L(0)$
of a single set of Virasoro generators $L(m)$, so open WZW string
theory on the Lie algebra $g$ is described by the zero mode $L_g(0)$ of the
affine-Sugawara construction [\ref{refBH},~\ref{refHa}-\ref{refSe},~\ref{refRev}] on $g$
\group{quantumstart}
\begin{gather}
H_g = L_g(0)=\onetwopi \int_{0}^{2\pi} d\xi \0b L^{ab}_{g}
:J_a(\xi,t)J_b(\xi,t):\,=L_g^{ab} \sum_{m\in\mathbb{Z}}:J_a(m)J_b(-m):
\label{hamiltonian}\\ [J_a(\xi,t), J_b(\eta,t)] = 2\pi
i\big{(}f_{ab}{}^c J_c(\xi,t) \d(\xi-\eta) +
G_{ab}\pl_\xi\d(\xi-\eta)\big{)} \\ \d(\xi\mp\eta) = \onetwopi
\sum_{m\in\mathbb{Z}} e^{im(\xi\mp\eta)}, \quad 0\le\xi,\eta\le
2\pi,\quad a,b,c=1,\ldots \mathrm{dim}g \ .
\end{gather}
\reseteqn
Here $f_{ab}{}^c$ and $G_{ab}$ are the structure constants
and metric of $g$
\begin{equation}
g=\oplus_I g^I,\quad f_{ab}{}^c=\oplus_I f_{ab}^I{}^c,\quad
G_{ab}=\oplus_I k_I \eta_{ab}^I
\end{equation}
where $I$ is the semisimplicity index and $k_I$ is the level of the
simple component $g^I$. The coefficient $L_g^{ab}$
\begin{equation}
L_g^{ab}=\oplus_I {\eta_I^{ab}\over 2k_I+Q_I}\;\stackrel{ \{k_I\} \gg
1}{\longrightarrow}\; L_{g,\infty}^{ab}\equiv {G^{ab}\over 2},\quad
G^{ab}=\oplus_I k_I^{-1} \eta^{ab}_I
\end{equation}
is the inverse inertia tensor of the affine-Sugawara construction on
$g$.

The Hamiltonian \eqref{hamiltonian} guarantees the chirality of the single 
non-abelian chiral current $J(\xi,t)$
\group{J(m)}
\begin{gather}
\pl_t A(\xi,t) = i [H_g,A(\xi,t)]\\ \pl_- J_a(\xi,t)=0,\quad
\pl_-\equiv \pl_t-\pl_\xi\\ J_a(\xi,t) = \sum_{m\in\mathbb{Z}}
J_a(m)e^{-im(t+\xi)} \\ [J_a(m), J_b(n)] = i f_{ab}{}^c J_c(m+n) + m
G_{ab} \d_{m+n,0} \label{modeJ}
\end{gather}
\reseteqn 
and Eq. \eqref{modeJ} is the affine Lie algebra
\cite{KM,BH,Rev} of $g$.

The affine-Sugawara Hamiltonian above is defined on the cylinder
$0\le\xi\le 2\pi$, but 
it is conventional to consider open string theory on the strip
$0\le\xi\le\pi$. Since the single chiral current is periodic when $\xi\to\xi+2\pi$,
the affine-Sugawara Hamiltonian may be rewritten in the open string picture
\begin{equation}
H_g=\onetwopi \int_{0}^{\pi} d\xi \0b L^{ab}_{g}
:\big{(}J_a(\xi,t)J_b(\xi,t)+J_a(-\xi,t)J_b(-\xi,t)\big{)}:
\label{openpicture}
\end{equation}
but where are the WZW branes? In what follows we answer this question
first in the classical formulation of the theory, returning to the
quantum theory in Sec.~7.

\vspace{-.07in}
\subsection{Classical Currents and Stress Tensors}

The classical version of the affine-Sugawara system on the cylinder 
is

\group{cylinder}
\begin{gather}
H_g = \int_{0}^{2\pi} d\xi \0b T_g(\xi,t)\\
T_g(\xi,t) = 
\onetwopi L_{g,\infty}^{ab} J_a(\xi,t) J_b(\xi,t)=\onefourpi G^{ab} J_a(\xi,t) J_b(\xi,t) \\ 
J_a(\xi+2\pi n,t)=J_a(\xi,t),\quad T_g(\xi+2\pi n,t)=T_g(\xi,t)\\
\begin{align} 
\{ J_a(\xi,t), J_b(\eta,t)\} =&\, 2\pi i\big{(}f_{ab}{}^c J_c(\xi,t)
\d(\xi-\eta) + G_{ab}\pl_\xi\d(\xi-\eta)\big{)}\\
\{T_g(\xi,t),T_g(\eta,t)\} =&\, i\big{(} T_g(\xi,t)+T_g(\eta,t)\big{)}
\pl_\xi\d(\xi-\eta)\\
\{T_g(\xi,t),J_a(\eta,t)\} =&\, i J_a(\xi,t) \pl_\xi\d(\xi-\eta)
\end{align} \\
0\le \xi,\eta \le 2\pi 
\end{gather}
\reseteqn
where $J_a(\xi,t)$ and 
$T_g(\xi,t)$ are the classical non-abelian chiral current and its classical stress tensor
respectively. In \eqref{cylinder} $\{\cdot,\cdot\}$ are Poisson brackets multiplied 
by a convenient extra factor of $i$.

To go to the open string picture, we decompose the single non-abelian cylinder current
and its single stress tensor into components on the strip as follows:

\group{decompose}
\begin{gather}
\begin{align}
J_a(\xi,t),\;\;0\le\xi\le 2\pi\quad&\rightarrow\quad J_a(\pm\xi,t),\;\; 0\le\xi\le \pi\\
T_g(\xi,t),\;\;0\le\xi\le 2\pi\quad&\rightarrow\quad T_g(\pm\xi,t),\;\;0\le\xi\le \pi
\end{align}\\
T_g(\pm\xi,t)=\onefourpi G^{ab} J_a(\pm\xi,t) J_b(\pm\xi,t)\\
\begin{align}
H_g = & \int_{0}^{\pi} d\xi \0b \big{(} T_g(\xi,t) + T_g(-\xi,t) \big{)}\\
= & \onefourpi \int_{0}^{\pi} d\xi \0b G^{ab}
\big{(}J_a(\xi,t)J_b(\xi,t)+J_a(-\xi,t)J_b(-\xi,t)\big{)}\label{decomposeham}\ .
\end{align}
\end{gather}
\reseteqn
The bracket algebra of the components on the strip $0\le\xi,\eta\le\pi$
then follows from (\ref{cylinder}c-e) 
\group{J-J}
\begin{align}
\{ J_a(\xi,t), J_b(\eta,t)\} =&\, 2\pi i\big{(}f_{ab}{}^c J_c(\xi,t)
\d(\xi-\eta) + G_{ab}\pl_\xi\d(\xi-\eta)\big{)} \\ \{J_a(\xi,t),
J_b(-\eta,t)\} =& \,2\pi i\big{(}f_{ab}{}^c J_c(\xi,t) \d(\xi+\eta) +
G_{ab}\pl_\xi\d(\xi+\eta)\big{)} \\ \{J_a(-\xi,t), J_b(-\eta,t)\} =&
\,2\pi i\big{(}f_{ab}{}^c J_c(-\xi,t) \d(\xi-\eta) -
G_{ab}\pl_\xi\d(\xi-\eta)\big{)} 
\end{align}
\reseteqn

\group{T-T}
\begin{align}
\{T_g(\xi,t),T_g(\eta,t)\} =& \,i\big{(} T_g(\xi,t)+T_g(\eta,t)\big{)}
\pl_\xi\d(\xi-\eta) \\ \{T_g(\xi,t),T_g(-\eta,t)\} =&\, i\big{(}
T_g(\xi,t)+T_g(-\eta,t)\big{)} \pl_\xi\d(\xi+\eta) \\
\{T_g(-\xi,t),T_g(-\eta,t)\} =& \,-i\big{(}
T_g(-\xi,t)+T_g(-\eta,t)\big{)} \pl_\xi\d(\xi-\eta)
\end{align}
\reseteqn

\group{J-T}
\begin{align}
\{T_g(\xi,t),J_a(\eta,t)\} =& \,i J_a(\xi,t) \pl_\xi\d(\xi-\eta) \\
\{T_g(\xi,t),J_a(-\eta,t)\} =& \,i J_a(\xi,t) \pl_\xi\d(\xi+\eta) \\
\{T_g(-\xi,t),J_a(\eta,t)\} =& \,-i J_a(-\xi,t) \pl_\xi\d(\xi+\eta) \\
\{T_g(-\xi,t),J_a(-\eta,t)\} =& \,-i J_a(-\xi,t) \pl_\xi\d(\xi-\eta)\ . 
\end{align}
\reseteqn
In (\ref{J-J}-\ref{J-T}), the delta function $\d(\xi-\eta)$
has support only at $\xi=\eta$, while the delta function $\d(\xi+\eta)$ has support
only at the strip boundary $\xi=\eta=0$ or $\pi$. We will refer to terms 
proportional to $\d(\xi-\eta)$ and $\d(\xi+\eta)$ as bulk and boundary 
terms respectively. In what follows, we interpret the form of the strip
current algebra \eqref{J-J}, remarking only that a similar interpretation applies to
\eqref{T-T} and \eqref{J-T}.

In discussing \eqref{J-J}, it is instructive to bear in mind the equal-time current algebra
of affine $(g\times g)$
\group{J-Jbar}
\begin{align}
\{ J_a(\xi,t), J_b(\eta,t)\} =& \,2\pi i\big{(}f_{ab}{}^c J_c(\xi,t)
\d(\xi-\eta) + G_{ab}\pl_\xi\d(\xi-\eta)\big{)} \\ 
\{J_a(\xi,t),\bar{J}_b(\eta,t)\} =& \,0 \\ 
\{\bar{J}_a(\xi,t), \bar{J}_b(\eta,t)\} =& \,2\pi i\big{(}f_{ab}{}^c \bar{J}_c(\xi,t) \d(\xi-\eta) -
G_{ab}\pl_\xi\d(\xi-\eta)\big{)} \\
0\le\xi,\eta\le 2\pi
\end{align}
\reseteqn
which holds in the ordinary WZW model. One sees that the brackets (\ref{J-J}a) and (\ref{J-J}c)
are locally isomorphic under $J(-\xi,t)\to\bar{J}(\xi,t)$ to the brackets (\ref{J-Jbar}a) and (\ref{J-Jbar}c),
but (\ref{J-J}b) tells us that this isomorphism fails at the boundary of the strip, where
\begin{equation}
\{J_a(\xi,t), J_b(-\eta,t)\}\ne 0 \quad \mathrm{at}\,\,\xi=\eta=0\,\mathrm{or}\,\pi
\end{equation}
due to the boundary term $\d(\xi+\eta)$. As we will see, this difference controls many
important aspects of open WZW theory. 

Although (\ref{J-J}b) is non-zero at the boundary, the strip current system (\ref{J-J}) is locally WZW
in the bulk, where the boundary terms do not contribute.

We may also interpret the boundary terms proportional to $\d(\xi+\eta)$ in 
(\ref{J-J}-\ref{J-T}) as due to the interaction of a non-abelian charge at $\xi$ (or $\eta$)
with a generalized non-abelian Dirichlet image charge at $-\eta$ (or $-\xi$).

\vspace{-.07in}
\subsection{Classical Dynamics}

In classical open WZW string theory, the time dependence of the fields is determined by
\begin{equation}
\pl_t A(\xi,t) = i \{H_g, A(\xi,t)\}
\label{plta}
\end{equation}
where $H_g$ is given in \eqref{decomposeham}.
For such computations, it is useful to record the following identities
\group{f}
\begin{gather}
\int_{0}^{\pi} d\eta \0b \d(\eta-\xi) f(\eta) = \left\{
\begin{array}{ll} \half f(0) & \mbox{if $\xi=0$} \\ f(\xi) & \mbox{if
$0<\xi<\pi$} \\\half f(\pi) & \mbox{if $\xi=\pi,$} \end{array} \right
. \\ \int_{0}^{\pi} d\eta \0b \d(\eta+\xi) f(\eta) = \left\{
\begin{array}{ll} \half f(0) & \mbox{if $\xi=0$} \\ 0 & \mbox{if
$0<\xi<\pi$} \\ \half f(\pi) & \mbox{if $\xi=\pi$} \end{array} 
\right .
\end{gather}
\reseteqn 
which will allow us to evaluate brackets of
integrated quantities. Then we find from \eqref{plta} and \eqref{f}
that the strip currents and stress tensors are chiral
\group{J}
\begin{gather}
\pl_- J_a(\xi,t) = \pl_+ J_a(-\xi,t)=0, \quad \pl_{\pm}\equiv\pl_t \pm
\pl_\xi\label{chiralcurrents}\\ J_a(\xi,t) = \sum_{m\in\mathbb{Z}}
J_a(m) e^{-im(t+\xi)},\quad J_a(-\xi,t) = \sum_{m\in\mathbb{Z}} J_a(m)
e^{-im(t-\xi)}\label{modealgebra}\\ \{J_a(m),J_b(n)\}= i f_{ab}{}^c
J_c(m+n) + m G_{ab} \d_{m+n,0} \\
\pl_- T(\xi,t) = \pl_+ T(-\xi,t)=0
\end{gather}
\reseteqn as expected from the quantum theory.

We may also consider the natural candidate for a momentum operator,
which we call the bulk momentum operator $P_g$:
\group{P}
\begin{gather}
\begin{align}
P_g(t) &\equiv \int_{0}^{\pi} d\xi \0b \big{(} T_g(\xi,t) -
T_g(-\xi,t) \big{)} \\ =& \onefourpi \int_{0}^{\pi} d\xi \0b
G^{ab} \big{(} J_a(\xi,t)J_b(\xi,t) - J_a(-\xi,t)J_b(-\xi,t) \big{)}
\end{align}\\
\pl_t P_g(t) = {G^{ab}\over 2\pi} \big{(} J_a(\pi,t)J_b(\pi,t) -
J_a(0,t)J_b(0,t)\big{)} \ .
\end{gather}
\reseteqn 
According to its name, $iP_g$ generates $\pl_\xi$ only in the
bulk. In the case of the currents, one finds for example
\begin{equation}
i\{P_g(t),J_a(\pm\xi,t)\}=\pl_\xi \left\{ \begin{array}{ll}
J_a(\pm\xi,t) & \mbox{if $0<\xi<\pi$}\\ 0 & \mbox{if $\xi=0,\pi$}
\end{array} \right .
\label{P-J}
\end{equation}
where we have used the equal-time current algebra \eqref{J-J} and the
relations \eqref{f}. In fact (as seen in this example) the correct
form of $\pl_\xi A$ for any $A$ can be obtained from the form of
$i\{P_g,A\}$ in the bulk
\begin{equation}
\pl_\xi A = i \{P_g(t),A\},\quad 0<\xi<\pi
\end{equation} 
by smoothly extending this form to the boundary. This observation will
be useful in the quantum theory.

\vspace{-.1in}
\newsection{Phase Space Realization of the Currents}

In what follows, we postulate the following phase space realization\footnote{Other
realizations of the equal time current algebra can be obtained by replacing 
$J_a(-\xi,t)$ on the left of (\ref{bowcock}b) by $J_a^\o(-\xi,t)=\o_a{}^b J_b(-\xi,t)$, where $\o$ 
is an element of the automorphism group of $g$. This replacement does not 
change the Hamiltonian and so leads to T-dual
forms of the theory, which will not be discussed in this paper.}of
the equal-time current algebra \eqref{J-J}
\group{bowcock}
\begin{gather}
\begin{align}
J_a(\xi,t)\equiv& 2\p e(x(\xi,t))_a{}^i p_i(B,\xi,t)+\half \pl_\xi
x^i(\xi,t) e(x(\xi,t))_i{}^b G_{ba}\\ J_a(-\xi,t)\equiv& 2\p
\eb(x(\xi,t))_a{}^i p_i(B,\xi,t)-\half \pl_\xi x^i(\xi,t)
\eb(x(\xi,t))_i{}^b G_{ba}
\end{align}\\
0\le\xi\le\pi \ .
\end{gather}
\reseteqn 
The quantities which appear in \eqref{bowcock} are defined
as follows
\group{def1}
\begin{gather}
e_i(T) \equiv e_i{}^a T_a = -i g^{-1}(T)\pl_i g(T), \quad \eb_i(T) 
\equiv \eb_i{}^a T_a = -i g(T)\pl_i g^{-1}(T) \label{viel}\\ e_i{}^a
 e_a{}^j = \eb_i{}^a \eb_a{}^j = \d_i{}^j \\ p_i(B,\xi,t)=
 p_i(\xi,t)+\onefourpi B_{ij}(x(\xi,t))\pl_\xi x^j(\xi,t) \label{pb}\\
 \pl_i B_{jk}(x) + \pl_j B_{ki}(x) + \pl_k B_{ij}(x) = -i\,
 \mathrm{Tr} \big{(}\,M(k,T) e_i(x,T) [e_j(x,T), e_k(x,T)] \,\big{)}
 \\ [T_a,T_b]=i f_{ab}{}^c T_c,\quad \mathrm{Tr}\big{(}M(k,T) T_a T_b
 \big{)} = G_{ab},\quad T_a=\oplus_I T_a^I,\quad M(k,T)\equiv \oplus_I
 {k_I\over y_I(T^I)}
\end{gather}
\reseteqn 
where $T_a$ is any matrix irrep of the Lie algebra $g$,
$g(T)$ are the group elements in irrep $T$ and $e_i{}^a$ and
$\eb_i{}^a$ are the left and right invariant vielbeins on the group
manifold, respectively. The antisymmetric tensor $B_{ij}$ is the $B$
field of the open WZW string and the data matrix $M$ stores
information about the Dynkin indices $y_I(T)$ of the simple factors
$T^I$.  In \eqref{def1}, the matrices $T$ and $g(T)$ are square, and
matrix multiplication is defined as
\begin{equation}
e_i{}^a (T_a)_\a{}^\b= -i g^{-1}(T)_\a{}^\c \pl_i g(T)_\c{}^\b,\quad
\a,\b,\c=1,\ldots,\mathrm{dim}\,T
\end{equation}
with a summation convention for repeated indices.

In Eq.~\eqref{bowcock}, our phase space realization of the currents
$J(\xi,t)$ and $J(-\xi,t)$, $0\le\xi\le\pi$ is the usual \cite{Bo} WZW
phase space realization of the left and the right mover currents
$J(\xi,t)$ and $\bar{J}(\xi,t)$, $0\le\xi\le 2\pi$. As we shall see,
this realization will allow us to require that the open WZW string has
the same bulk dynamics ($0<\xi<\pi$) as the WZW model.

Indeed, Eq.~\eqref{bowcock} gives the phase space form of the Hamiltonian
\group{ham}
\begin{gather}
H_g = \int_{0}^{\pi} d\xi \0b \mathcal{H}_g\\
\mathcal{H}_g=\frac{G^{ab}}{4\pi}(J_a(\xi)J_b(\xi)+J_a(-\xi)J_b(-\xi))=2\pi
G^{ij} p_i(B) p_j(B) + \frac{1}{8\p} \pl_\xi x^i \pl_\xi x^j G_{ij} \\
G_{ij} = e_i{}^a e_j{}^b G_{ab} = \eb_i{}^a \eb_j{}^b G_{ab}
\end{gather}
\reseteqn whose density $\mathcal{H}_g$ is the same as that of the WZW
model. The bulk momentum operator
\begin{equation}
P_g(t)=\int_0^\pi d\xi \0b \pl_\xi x^i(\xi,t) p_i(\xi,t)
\end{equation}
also has the usual WZW momentum density.

Moreover, the equal-time current algebra \eqref{J-J} and its phase
space realization \eqref{bowcock} imply a system of constraints on
the phase space variables, which will allow us to obtain the phase
space bracket algebra itself. As a first example of these constraints, we note
two equivalent forms of the boundary conditions
\group{consistency}
\begin{gather}
J_a(0,t) = J_a(-0,t), \quad J_a(\pi,t) = J_a (-\pi,t)
\label{consistency1}\quad \mathrm{or}\\ 4\pi(\eb(\xi,t)_a{}^i -
e(\xi,t)_a{}^i) p_i(B,\xi,t) = \pl_\xi x^i(\xi,t) (\eb(\xi,t)_i{}^b +
e(\xi,t)_i{}^b) G_{ba} \quad {\rm at} \,\, \xi=0,\pi 
\end{gather}
\reseteqn which follow from the $2\pi$ periodicity of the cylinder current. 
We will see below that these boundary conditions are
equivalent to the generalized Dirichlet boundary conditions of
Ref.~\cite{AS}.

\vspace{-.1in}
\newsection{Phase Space Algebra: First Results}

\vspace{-.07in}
\subsection{The Inverse Relations and $\pl_\xi g$}

In this section, we will use the equal-time current algebra
\eqref{J-J} and its phase space realization \eqref{bowcock} to begin
our analysis of the phase space brackets of open WZW theory. 

In preparation for this analysis, we will need the inverse relations
 \group{inverse}
\begin{gather}
\pl_\xi x^i(\xi,t) = J_a(\xi,t) G^{ab} e(\xi,t)_b{}^i - J_a(-\xi,t)
G^{ab} \eb(\xi,t)_b{}^i \\ p_i(B,\xi,t) = \onefourpi \big{(}
e(\xi,t)_i{}^a J_a(\xi,t) + \eb(\xi,t)_i{}^a J_a(-\xi,t)\big{)}
\end{gather}
\reseteqn which follow from \eqref{bowcock}.

As a first application of the inverse relations, we note the following form
of $\pl_\xi g$
\group{partial g}
\begin{gather}
\pl_\xi g(T,\xi,t) = \pl_\xi x^i \pl_i g(T,\xi,t) = i \big{(}
g(T,\xi,t) J(T,\xi,t) + J(T,-\xi,t) g(T,\xi,t)\big{)} \\ J(T,\pm
\xi,t) \equiv J_a(\pm \xi,t) G^{ab} T_b
\end{gather}
\reseteqn which is obtained by chain rule from \eqref{inverse} and \eqref{def1}. 
The result \eqref{partial g}
is the usual form (with $\bar{J}(\xi)\to J(-\xi)$) of $\pl_\xi g$ in
the WZW model.

\vspace{-.07in}
\subsection{Bracket of $J$ with $x$}

Using the equal-time current algebra \eqref{J-J} and the inverse
relations \eqref{inverse}, we may derive a differential equation for
the bracket of $J(\xi,t)$ with $x$,
\begin{align}
\nonumber \pl_\eta\{ J_a(\xi,t), x^i(\eta,t) \} =& \{ J_a(\xi,t),
\pl_\eta x^i(\eta,t) \}\\ \nonumber =& \{ J_a(\xi), J_b(\eta) G^{bc}
e(\eta)_c{}^i - J_b(-\eta) G^{bc} \eb(\eta)_c{}^i \} \\ \nonumber =&
2\pi i \big{(} f_{ab}{}^c J_c(\xi) \d(\xi-\eta) + G_{ab}\pl_\xi
\d(\xi-\eta)\big{)} G^{bd} e(\eta)_d{}^i \\ \nonumber &-2\pi i \big{(}
f_{ab}{}^c J_c(\xi) \d(\xi+\eta) + G_{ab}\pl_\xi \d(\xi+\eta)\big{)}
G^{bd} \eb(\eta)_d{}^i \\ &+\{ J_a(\xi), x^j(\eta)\}\big{(} \pl_j
e(\eta)_b{}^i G^{bc} J_c(\eta) - \pl_j \eb(\eta)_b{}^iG^{bc}
J_c(-\eta) \big{)}\ .
\label{Jx-eq}
\end{align}
The same equation with $\xi\to -\xi$ is obtained for the bracket of
$J(-\xi,t)$ with $x(\eta,t)$. We have checked that the ordinary WZW
bracket
\begin{equation}
\{J_a(\xi,t), x^i(\eta,t) \}_\mathrm{WZW}= -2\pi i
e(\eta,t)_a{}^i\d(\xi-\eta)
\end{equation}
is not a solution to \eqref{Jx-eq}. A particular solution to these
equations is
\group{J-x}
\begin{gather}
\{ J_a(\xi,t), x^i(\eta,t) \} = -2\pi i \big{(}
e(\eta,t)_a{}^i\d(\xi-\eta) + \eb(\eta,t)_a{}^i \d(\xi+\eta) \big{)}
\label{J-x1}\\ \{ J_a(-\xi,t), x^i(\eta,t) \} = -2\pi i \big{(}
\eb(\eta,t)_a{}^i\d(\xi-\eta) + e(\eta,t)_a{}^i \d(\xi+\eta) 
\big{)} \ .
\end{gather}
\reseteqn Among these terms only the bulk terms (proportional to
$\d(\xi-\eta)$) are present in the ordinary WZW model, while the terms
proportional to $\d(\xi+\eta)$ are boundary terms which represent
non-abelian generalized Dirichlet images.

In finding these solutions from \eqref{Jx-eq}, we used the relation
\begin{equation}
J_a(\xi,t) \d(\xi+\eta) = J_a(-\eta,t) \d(\xi+\eta),\quad
0\le\xi,\eta\le\pi
\end{equation}
which holds because the cylinder current is $2\pi$ periodic. Verification of
these solutions requires considerable algebra: In particular, one
needs the explicit form \eqref{bowcock} of $J(\pm\xi,t)$ and the
Cartan-Maurer identities, as well as the identities
\begin{equation}
\eb_i{}^a = - e_i{}^b \Omega_b{}^a, \quad \partial_i \Omega_a{}^b =
f_{ac}{}^d e_i{}^c \Omega_d{}^b,\quad \Omega_a{}^c \Omega_b{}^d
G_{cd}=G_{ab}
\label{adjoint}
\end{equation}
where $\Omega$ is the adjoint action.

The general solution to the differential equations for these brackets
is obtained by adding to the particular solution (\ref{J-x}a-b) the
terms
\group{general}
\begin{gather}
\delta \{ J_a(\xi,t), x^i(\eta,t) \} = f(\xi,t) B(\eta,t)_a{}^i\\
\delta \{ J_a(-\xi,t), x^i(\eta,t) \} = h(\xi,t) B(\eta,t)_a{}^i \\
f(0,t)=h(0,t),\quad f(\pi,t)=h(\pi,t)\label{bcforgh}\\ \pl_\eta
B(\eta,t)_a{}^i = B(\eta,t)_a{}^j \big{(} \pl_j e(\eta,t)_b{}^i G^{bc}
J_c(\eta,t) - \pl_j \eb(\eta,t)_b{}^iG^{bc} J_c(-\eta,t) \big{)}
\end{gather}
\reseteqn
where $f(\xi,t)$ and $h(\xi,t)$ are arbitrary except for the boundary
conditions in \eqref{bcforgh}.

Following our program, we set these terms to zero in order to maintain
the ordinary WZW term $-2\pi i e_a{}^i \d(\xi-\eta)$ in the bulk.

Note that the solutions \eqref{J-x} are consistent with the
generalized Dirichlet boundary conditions \eqref{consistency1}
because
\begin{equation}
\d(-\eta)=\d(\eta),\quad \d(\pi-\eta)=\d(\pi+\eta).
\end{equation} 
Moreover, the boundary terms proportional to $\d(\xi+\eta)$ are
necessary for this consistency.

By chain rule from the brackets of $J(\pm\xi,t)$ with $x$, we also
find the brackets of the currents with the group elements
\group{J-g}
\begin{gather}
\{J_a(\xi,t), g(T,\eta,t) \} = 2\pi \big{(} \d(\xi-\eta) g(T,\eta,t)
T_a - \d(\xi+\eta) T_a g(T,\eta,t) \big{)} \\ \{J_a(-\xi,t),
g(T,\eta,t) \} = 2\pi \big{(} -\d(\xi-\eta) T_a g(T,\eta,t) +
\d(\xi+\eta) g(T,\eta,t) T_a \big{)}\ .
\end{gather}
\reseteqn As seen for $\{J(\pm\xi),x\}$, these brackets are also
consistent with the generalized Dirichlet boundary conditions
\eqref{consistency1}.  Here the first relation shows a right rotation
in the bulk and a left rotation at the boundary, and vice-versa for
the second relation.  The brackets \eqref{J-g} will be central in the
quantization of the open WZW string in Sec.~7.

As an application of Eqs.~\eqref{J-g} and \eqref{f}, we may compute
the action of the bulk momentum operator \eqref{P} on the group
elements
\begin{gather}
i\{P_g(t),g(T,\xi,t)\}=\left\{ \begin{array}{ll} i \big{(} g(T,\xi,t)
J(T,\xi,t) + J(T,-\xi,t) g(T,\xi,t)\big{)} & \mbox{if $0<\xi<\pi$}\\ 0
& \mbox{if $\xi=0,\pi$.} \end{array} \right .
\label{P-g}
\end{gather}
\reseteqn Comparing this result with $\pl_\xi g$ in
Eq.~\eqref{partial g}, we see that the correct form of $\pl_\xi g$
can be obtained by smoothly extending to the boundary the form of
$i\{P,g\}$ obtained in the bulk. The same conclusion was obtained 
for the currents in \eqref{P-J}.

\vspace{-.1in}
\newsection{Coordinate Space}

\vspace{-.07in}
\subsection{Equations of Motion}

We postpone further study of the phase space bracket algebra, because we
already know enough to make the transition to coordinate space.

Consider the computation
\group{time-x}
\begin{gather}
\begin{align}
\nonumber \partial_t x^i (\xi,t) =& i\{H_g,x^i(\xi,t)\} \\ \nonumber
=& G^{ab} \int_{0}^{\pi} d\eta \0b \big{[} \big{(}
e(\xi)_a{}^iJ_b(\eta)+ \eb(\xi)_a{}^iJ_b(-\eta)\big{)} \d(\eta-\xi) \\
\nonumber &\quad \quad \quad \quad \quad \quad +\big{(}
\eb(\xi)_a{}^iJ_b(\eta)+ e(\xi)_a{}^iJ_b(-\eta)\big{)}
\d(\eta+\xi)\big{]} \\ \nonumber =& G^{ab} \big{(}
e(\xi)_a{}^iJ_b(\xi)+ \eb(\xi)_a{}^iJ_b(-\xi)\big{)} \\ =& 4\pi
G^{ij}(\xi,t) p_j(B,\xi,t)
\end{align} \\
p_i(B,\xi,t)=\onefourpi G_{ij}(\xi,t)\pl_t x^j(\xi,t) \label{px}
\end{gather}
\reseteqn where we have used the results \eqref{J-x}, \eqref{f} and
\eqref{consistency1}.  This is the usual local relation between
$\pl_t x$ and $p$ in the WZW model.  Then the relations
\eqref{time-x}, \eqref{inverse} and the chain rule give the
time-derivative of the group elements in terms of the currents. We
record this result along with the space-derivative of $g$ derived in
\eqref{partial g}:
\group{time-g}
\begin{gather}
\pl_t g(T,\xi,t) = i \big{(} g(T,\xi,t)J(T,\xi,t) - J(T,-\xi,t)
g(T,\xi,t) \big{)} \label{pltg}\\ \pl_\xi g(T,\xi,t) = i \big{(}
g(T,\xi,t)J(T,\xi,t) + J(T,-\xi,t) g(T,\xi,t) \big{)} \\ \pl_+
g(T,\xi,t) = 2i g(T,\xi,t) J(T,\xi,t), \quad \pl_- g(T,\xi,t) = -2i
J(T,-\xi,t) g(T,\xi,t)\label{classicalpm} \ .
\end{gather}
\reseteqn The same result for $\pl_t g$ can be obtained from
\eqref{plta} and the brackets $\{J(\pm\xi),g\}$.  From the phase
space realization of the currents and the relation between $p$ and
$\pl_t x$, we obtain the coordinate space form of $J$:
\group{currents}
\begin{gather}
J_a(\xi,t)= \half \pl_+ x^i(\xi,t) e(\xi,t)_i{}^b G_{ba}, \quad
J_a(-\xi,t)= \half \pl_- x^i(\xi,t) \eb(\xi,t)_i{}^b G_{ba} \\
J(T,\xi,t)= -\frac{i}{2} g^{-1}(T,\xi,t) \pl_+ g(T,\xi,t), \quad
J(T,-\xi,t)= -\frac{i}{2} g(T,\xi,t) \pl_- g^{-1}(T,\xi,t)
\label{currents2}\\ \pl_-(g^{-1}(T,\xi,t) \pl_+
g(T,\xi,t))=\pl_+(g(T,\xi,t) \pl_- g^{-1}(T,\xi,t))=0
\label{currents3}\ .
\end{gather}
\reseteqn The results in \eqref{currents2} are equivalent to
Eq.~\eqref{classicalpm}. The relations in \eqref{currents3} 
(which are the usual WZW equations of motion) follow from 
\eqref{currents2} and the
chirality \eqref{chiralcurrents} of the currents.

\vspace{-.07in}
\subsection{Generalized Dirichlet Boundary Conditions}

Using \eqref{px}, we see that the boundary conditions
\eqref{consistency} can now be written in the following three
equivalent forms:
\group{bc}
\begin{gather}
J(0,t) = J(-0,t), \quad J(\pi,t) = J(-\pi,t)\quad\mathrm{or}\label{bc1}\\ \pl_t
x^i(\xi,t) (\eb(\xi,t)_i{}^a - e(\xi,t)_i{}^a) = \pl_\xi x^i(\xi,t)
(\eb(\xi,t)_i{}^a + e(\xi,t)_i{}^a)\quad\mathrm{or}\label{bc2}\\
g^{-1}(T,\xi,t) \pl_+ g(T,\xi,t) = g(T,\xi,t) \pl_- g^{-1}(T,\xi,t)
\quad {\rm at}\,\, \xi=0,\pi \label{bc3}
\end{gather}
\reseteqn where \eqref{bc3} is the generalized Dirichlet boundary
condition of Ref.~\cite{AS}. The result \eqref{bc} shows that our open WZW 
strings end on WZW branes.

\vspace{-.07in}
\subsection{The Bulk Lagrange Density}

We define a bulk Lagrange density $\mathcal{L}_g$ for the open WZW string
by the usual Legendre transformation
\group{action}
\begin{gather}
\sl_g \equiv \pl_t x^i p_i - \mathcal{H}_g =
\frac{1}{8\pi} (G_{ij}+B_{ij}) \pl_+ x^i \pl_-
x^j,\quad 0<\xi<\pi \label{sigma-model}\\
\frac{1}{8\pi} G_{ij} \pl_+ x^i \pl_- x^j=
-\frac{1}{8\pi} \mathrm{Tr}\big{(}M(k,T) g^{-1}(T)\pl_+ g(T)
g^{-1}(T)\pl_- g(T)\big{)}\ .
\end{gather}
\reseteqn 
The bulk density $\sl_g$ in \eqref{sigma-model} is the sigma model form of the
usual WZW Lagrange density, and $M$ in (\ref{action}b) is the data matrix 
defined in \eqref{def1}. The local equations of motion
of this density agree in the bulk with our Hamiltonian equations of 
motion \eqref{currents3} so, again, our Hamiltonian formulation is 
locally WZW in the bulk.

\vspace{-.1in}
\newsection{New Equal-Time Non-Commutative Geometry} 

In this section, we continue our phase space construction to find the
equal-time coordinate brackets
\begin{equation}
\D^{ij}(\xi,\eta,t)\equiv \{ x^i(\xi,t),x^j(\eta,t) \} \ .
\label{x-x}
\end{equation}
In particular, we may use the inverse relations \eqref{inverse} and
the equal-time brackets \eqref{J-x} of $J(\pm\xi)$ with $x$ to find
the differential equation
\group{xx-eq}
\begin{gather} 
\begin{align} \nn 
\pl_\eta \D^{ij}(\xi,\eta,t) =& \{ x^i(\xi,t), \pl_\eta x^j(\eta,t) \}
                           \\ \nn =& \{ x^i(\xi), J_a(\eta) G^{ab}
                           e(\eta)_b{}^i - J_a(-\eta) G^{ab}
                           \eb(\eta)_b{}^i \}\\ =& 2\pi i \Psi^{ij}
                           (\xi,\eta) \d(\xi+\eta) + \D^{ik}(\xi,\eta)
                           \L(\eta)_k{}^j
\end{align} \\
\Psi^{ij}(\xi,\eta,t) \equiv \eb(\xi,t)_a{}^i G^{ab} e_b{}^j(\eta,t) -
e(\xi,t)_a{}^i G^{ab} \eb_b{}^j(\eta,t)=-\Psi^{ji}(\eta,\xi,t) \\
\L(\xi,t)_i{}^j \equiv J_a(\xi,t) G^{ab} \pl_i e_b{}^j(\xi,t) -
J_a(-\xi,t) G^{ab} \pl_i \eb_b{}^j(\xi,t)
\end{gather}
\reseteqn and a similar equation for $\pl_\xi \D$.

In a matrix notation, these two equations read
\group{xx-eqmatrix}
\begin{gather}
\pl_\eta \D(\xi,\eta,t) = 2 \pi i \Psi(\xi,\eta,t) \d(\xi+\eta)+
\D(\xi,\eta,t) \L(\eta,t) \\ \pl_\xi \D(\xi,\eta,t) = 2\pi i
\Psi(\xi,\eta,t) \d(\xi+\eta) + \L^\mathrm{T}(\xi,t) \D(\xi,\eta,t)
\label{xx-eqmatrixb}\\
\Psi^\mathrm{T}(\eta,\xi)=-\Psi(\xi,\eta)
\end{gather}
\reseteqn where $\mathrm{T}$ is matrix transpose. The integrability
condition for this system is
\group{int}
\begin{gather}
\pl_\xi \pl_\eta \D(\xi,\eta,t) = \pl_\eta \pl_\xi
\D(\xi,\eta,t)\quad\mathrm{iff} \\ \big{(}\pl_\eta \Psi(\xi,\eta,t) -
\Psi(\xi,\eta,t) \L(\eta,t) \big{)} \d(\xi+\eta) = \big{(}\pl_\xi
\Psi(\xi,\eta,t) - \L^\mathrm{T}(\xi,t) \Psi(\xi,\eta,t) \big{)}
\d(\xi+\eta) \ .
\end{gather}
\reseteqn Using the definitions of $\Psi$ and $\L$ in
Eq.~\eqref{xx-eq}, we find after some algebra that the integrability
condition is satisfied. The inhomogeneous terms in \eqref{xx-eqmatrix} 
are boundary terms associated to the interaction between a non-abelian 
charge at $\xi$ (or $\eta$) and a generalized non-abelian Dirichlet image charge 
at $-\eta$ (or $-\xi$). Note that, because of the inhomogeneous terms,
the WZW bracket $\D_\mathrm{WZW}(\xi,\eta,t)=0$ is not a solution of
the equations \eqref{xx-eqmatrix}.

By standard methods, one finds the solution for the coordinate
brackets
\begin{align} \nonumber
\D(\xi,\eta,t) =& U^\mathrm{T}(\xi,t) \D(0,0,t) U(\eta,t) \\ \nonumber
&+ \pi i \int_0^\eta d\eta^\prime \0b \big{(} \Psi(\xi,\eta^\prime,t)
\d(\xi+\eta^\prime) +
U^\mathrm{T}(\xi,t)\Psi(0,\eta^\prime,t)\d(\eta^\prime) \big{)}
U^{-1}(\eta^\prime,t) U(\eta,t) \\ &+ \pi i U^\mathrm{T}(\xi,t)
\int_0^\xi d\xi^\prime \0b U^{-1\,\mathrm{T}}(\xi^\prime,t) \big{(}
\Psi(\xi^\prime,\eta,t) \d(\xi^\prime+\eta) +
\Psi(\xi^\prime,0,t)\d(\xi^\prime) U(\eta,t) \big{)}
\label{xx-sol}
\end{align}
where $\D(0,0,t)$ is so far undetermined and we have introduced the
ordered product $U$ of $\L$
\group{U}
\begin{gather}
\pl_\xi U(\xi,t) = U(\xi,t) \L(\xi,t), \quad \pl_\xi
U^\mathrm{T}(\xi,t) = \L^\mathrm{T}(\xi,t) U^\mathrm{T}(\xi,t) \\
U(0,t)=U^\mathrm{T}(0,t)=1 \ .
\end{gather}
\reseteqn To check, for example, that \eqref{xx-sol} solves
\eqref{xx-eqmatrixb}, one needs the identity
 \begin{align}\nn
\pl_\xi \int_0^\eta d\eta^\prime \0b \Psi(\xi,\eta^\prime,t)
U^{-1}(\eta^\prime,t) \d(\xi+\eta^\prime) =& \L^\mathrm{T}(\xi,t)
\int_0^\eta d\eta^\prime \0b \Psi(\xi,\eta^\prime,t)
U^{-1}(\eta^\prime,t) \d(\xi+\eta^\prime)\\ &+ \Psi(\xi,\eta,t)
U^{-1}(\eta,t) \d(\xi+\eta) - \Psi(\xi,0,t) \d(\xi)
\label{identity}
\end{align}
which itself follows from the integrability condition \eqref{int}. If
we assume that the coordinate brackets are matrix antisymmetric at
$\xi=\eta=0$
\begin{equation}
\D^\mathrm{T}(0,0,t) = - \D(0,0,t)
\end{equation}
then \eqref{xx-sol} shows the correct antisymmetry of the coordinate 
brackets for all $\xi,\eta$
\begin{equation}
\D^\mathrm{T}(\eta,\xi,t) = - \D(\xi,\eta,t)
\end{equation}
because $\Psi$ has the same antisymmetry. We may also evaluate
\eqref{xx-sol} explicitly, with the result
\begin{equation}
\D(\xi,\eta,t)= \left\{\begin{array}{ll} \D(0,0,t), & \mbox{if
$\xi=\eta=0$} \\
U^\mathrm{T}(\pi,t)\big{(}\D(0,0,t)+i\pi\Psi(0,0,t)\big{)}U(\pi,t)+i\pi\Psi(\pi,\pi,t)
& \mbox{if $\xi=\eta=\pi$} \\
U^\mathrm{T}(\xi,t)\big{(}\D(0,0,t)+i\pi\Psi(0,0,t)\big{)}U(\eta,t) &
\mbox{otherwise.} \end{array} \right .
\end{equation}
This expression is suitable for the computation of $\xi$-derivatives
in the bulk, but one must return to \eqref{xx-sol} to compute
$\xi$-derivatives at the boundary.

The preferred choice of $\D(0,0,t)$ is
\begin{equation}
\D(0,0,t)=-i\pi\Psi(0,0,t)
\end{equation}
because in this case, as in the WZW model, the coordinate brackets
vanish in the bulk:
\group{xx-final}
\begin{gather}
\begin{align} \nonumber
\D(\xi,\eta,t) =& -i\pi U^\mathrm{T}(\xi,t) \Psi(0,0,t) U(\eta,t) \\
\nonumber &+ \pi i \int_0^\eta d\eta^\prime \0b \big{(}
\Psi(\xi,\eta^\prime,t) \d(\xi+\eta^\prime) +
U^\mathrm{T}(\xi,t)\Psi(0,\eta^\prime,t)\d(\eta^\prime) \big{)}
U^{-1}(\eta^\prime,t) U(\eta,t) \\ &+ \pi i U^\mathrm{T}(\xi,t)
\int_0^\xi d\xi^\prime \0b U^{-1\,\mathrm{T}}(\xi^\prime,t) \big{(}
\Psi(\xi^\prime,\eta,t) \d(\xi^\prime+\eta) +
\Psi(\xi^\prime,0,t)\d(\xi^\prime) U(\eta,t) \big{)}
\end{align}\\
\{x^i(\xi,t),x^j(\eta,t)\}=\D^{ij}(\xi,\eta,t)=
\left\{\begin{array}{ll} -i\pi\Psi^{ij}(0,0,t) & \mbox{if
$\xi=\eta=0$} \\ +i\pi\Psi^{ij}(\pi,\pi,t) & \mbox{if $\xi=\eta=\pi$}
\\ 0 & \mbox{otherwise} \end{array} \right .\\ \Psi^{ij}(\xi,\eta,t) =
\eb(\xi,t)_a{}^i G^{ab} e_b{}^j(\eta,t) - e(\xi,t)_a{}^i G^{ab}
\eb_b{}^j(\eta,t)\ .
\end{gather}
\reseteqn Eq.~\eqref{xx-final}, which shows the new non-commutative
geometry of open WZW strings, is a central result of this paper.

We emphasize that this new non-commutative geometry is an
intrinsically non-abelian effect, because the antisymmetric tensor
$\Psi$ vanishes in the abelian limit:
\group{1abelianlimit}
\begin{gather}
f_{ab}{}^c=0,\quad e_i{}^a=\d_i{}^a,\quad \eb_i{}^a=-\d_i{}^a\\
\Psi^{ij}=0,\quad \{x^i(\xi,t),x^j(\eta,t)\}=0 \ .
\end{gather}
\reseteqn In this abelian limit, our coordinates become Dirichlet and no
non-commutativity is expected for Dirichlet coordinates. Our new
non-abelian non-commutativity is therefore unrelated in any simple
sense to the standard [\ref{refCDS}-\ref{refSW}] non-commutativity found for
Neumann coordinates in the presence of a constant magnetic field.  At
$B=0$, other aspects of the abelian limit are noted in App.~A.

Using \eqref{xx-final}, \eqref{viel} and the chain rule, we have also
computed the brackets of the group elements among themselves
\group{g-g}
\begin{gather}
\{g(\xi,t)_\a{}^\b,g(\eta,t)_\c{}^\d\}\hspace{-.05in}=\hspace{-.05in}\left\{\begin{array}{ll}
\hspace{-.1in} i\pi(g(0,t)T_a)_\a{}^\b G^{ac}
(\Omega^{-1}(0,t)-\Omega(0,t))_c{}^b (g(0,t)T_b)_\c{}^\d
&\hspace{-.13in}\mbox{if $\xi=\eta=0$} \\
\hspace{-.1in} -i\pi(g(\pi,t)T_a)_\a{}^\b G^{ac}
(\Omega^{-1}(\pi,t)-\Omega(\pi,t))_c{}^b (g(\pi,t)T_b)_\c{}^\d
&\hspace{-.13in} \mbox{if $\xi=\eta=\pi$} \\
\hspace{-.1in}0 &\hspace{-.13in}\mbox{otherwise} \end{array} \right
.\\ \a,\b,\c,\d=1,\ldots\mathrm{dim}T
\end{gather}
\reseteqn
where $g\equiv g(T)$ and $\Omega$ is the adjoint action in \eqref{adjoint}. 

All other phase space brackets can now be straightforwardly obtained
from the known brackets $\{J(\pm \xi),x\}$ and $\{x,x\}$,
without solving any further differential equations. In
particular, App.~B gives the explicit form of the brackets:
\group{otherbrackets}
\begin{gather}
\{x^i(\xi,t),p_j(B,\eta,t)\},\quad \{x^i(\xi,t),p_j(\eta,t)\},\quad
\{J_a(\pm\xi,t),p_j(B,\eta,t)\}\\ \{p_i(B,\xi,t),p_j(B,\eta,t)\},\quad
\{p_i(\xi,t),p_j(\eta,t)\}\ .
\end{gather}
\reseteqn  The last
set of brackets $\{p,p\}$ is again non-zero only at the boundary, and
this effect again vanishes in the abelian limit.

\vspace{-.1in}

\newsection{The Conformal Field Theory of Open WZW Strings}

\vspace{-.07in}
\subsection{The Quantum Vertex Operators $g(T)$}

The Hamiltonian of the quantum theory of open WZW strings was given in
Sec.~1 
\group{quantumham}
\begin{gather}
\begin{align}
H_g=L_g(0)=&\onetwopi \int_{0}^{\pi} d\xi \0b L^{ab}_{g}
:J_a(\xi,t)J_b(\xi,t)+J_a(-\xi,t)J_b(-\xi,t):\\ =&L_g^{ab}
\sum_{m\in\mathbb{Z}} :J_a(m) J_b(-m):\\ =&L_g^{ab} \big{(} J_a(0)
J_b(0) + 2 \sum_{m=1}^\infty J_a(-m) J_b(m) \big{)}
\end{align}\\
:J_a(m) J_b(n):\,\equiv \theta(m\geq 0) J_b(n) J_a(m) + \theta(m<0)
J_a(m) J_b(n)\\ [J_a(m), J_b(n)] = i f_{ab}{}^c J_c(m+n) + m G_{ab}
\d_{m+n,0}\\ \pl_t A(\xi,t)=i[A(\xi,t),t]
\end{gather}
\reseteqn where the current modes $J_a(m)$ generate the affine Lie
algebra \cite{KM,BH,Rev} of $g$ and $L_g(0)$ is the zero mode of the
affine-Sugawara construction \cite{BH,Ha,DF,KZ,Se,Rev} on $g$.

Using the classical theory as a guide, and in particular \eqref{J-g}, 
we may now augment the quantum
system \eqref{quantumham} with the equal-time commutators
\group{J-gq}
\begin{gather}
\begin{align}
[J_a(\xi,t), g(T,\eta,t) ] =& 2\pi \big{(} g(T,\eta,t) T_a
\d(\xi-\eta) - T_a g(T,\eta,t) \d(\xi+\eta) \big{)} \\ [J_a(-\xi,t),
g(T,\eta,t) ] =& 2\pi \big{(} - T_a g(T,\eta,t) \d(\xi-\eta) +
g(T,\eta,t) T_a \d(\xi+\eta) \big{)}
\end{align}
\end{gather}
\reseteqn of the currents with the open string quantum vertex operators
$g(T)$. We emphasize that, as in the classical theory, these commutators 
are consistent with the generalized Dirichlet boundary conditions \eqref{bc1}.
Moreover, using the mode expansions \eqref{modealgebra} of the
currents, we find that the combined system \eqref{J-gq} is equivalent
to the algebra
\begin{equation}
[J_a(m), g(T,\xi,t)] = g(T,\xi,t) T_a e^{im(t+\xi)} - T_a g(T,\xi,t)
e^{im(t-\xi)}
\label{modeJ-gq}
\end{equation}
of the current modes with the vertex operators. We have checked that
the commutator \eqref{modeJ-gq} satisfies the $J,J,g$ Jacobi
identity, as it must. The Hamiltonian \eqref{quantumham} and the 
commutators \eqref{J-gq} or \eqref{modeJ-gq} will allow us to find
the analogues of the Knizhnik-Zamolodchikov equations for  
the conformal field theory of open WZW strings. 

\vspace{-.07in}
\subsection{Time Dependence}

Towards this goal, we first follow standard methods (see e.g. Halpern and Obers
\cite{HO}) to obtain the time differential equation for the open string vertex
operators.  The result is \group{voe1}
\begin{gather}
\pl_t g(T,\xi,t)=i[H_g,g(T,\xi,t)]\\
\begin{align}\nn
\pl_t g(T,\xi,t) =& 2i L_g^{ab} : g(T,\xi,t) J_a(\xi,t) T_b -
J_a(-\xi,t) T_b g(T,\xi,t):\\ &-2i L_g^{ab} T_a g(T,\xi,t) T_b + 2i
\D_g(T) g(T,\xi,t)\label{voe1result}\ .
\end{align}
\end{gather}
\reseteqn Here the normal ordering is defined as
\group{voe2}
\begin{gather}
: g(T,\xi,t) J_a(\pm\xi,t) : \, \equiv J_a^{(-)}(\pm\xi,t) g(T,\xi,t)
+ g(T,\xi,t) J_a^{(+)}(\pm\xi,t)\label{no}\\
J_a^{(+)}(\pm\xi,t)\equiv\sum_{m\geq 0} J_a(m) e^{-im(t\pm\xi)}\quad
J_a^{(-)}(\pm\xi,t)\equiv\sum_{m<0} J_a(m) e^{-im(t\pm\xi)}\\
J_a^{(+)}(\pm\xi,t)+J_a^{(-)}(\pm\xi,t)=J_a(\pm\xi,t)
\end{gather}
\reseteqn and $\D_g(T)$ is the conformal weight of irrep $T$ under the
affine Sugawara-construction on $g$
\begin{equation}
L_g^{ab} T_a T_b=\D_g(T) \one \ .
\end{equation}
In Eq.~\eqref{voe1result}, the normal-ordered terms have the same
form as the classical result Eq.~\eqref{pltg}, while the extra terms
are quantum effects from the normal ordering.

In order to study correlators of the open string vertex operators, we introduce the usual affine ground
state 
\group{vacuum}
\begin{gather}
J_a(m\geq 0) |0\rangle = \langle 0| J_a(m\leq 0) = 0 \\
J_a^{(+)}(\pm\xi,t) |0\rangle = \langle 0| J_a^{(-)}(\pm\xi,t) = 0 \ .
\end{gather}
\reseteqn This gives immediately the $g$-global Ward identities
\group{ward}
\begin{gather}
A(T,\xi,t)_\a{}^\b\equiv
A(T,\xi,t)_{\a_1\ldots\a_n}^{\;\;\;\b_1\ldots\b_n}\equiv \langle
g(T^1,\xi_1,t_1)_{\a_1}{}^{\b_1}\ldots
g(T^n,\xi_n,t_n)_{\a_n}{}^{\b_n} \rangle\label{correlator}\\
\a_i,\b_i=1,\ldots\mathrm{dim}\,T^i\\ A(T,\xi,t) \equiv \langle
g(T^1,\xi_1,t_1)\ldots g(T^n,\xi_n,t_n)\rangle,\quad q_a\equiv
\sum_{i=1}^n T_a^i\\ \langle [J_a(0), g(T^1,\xi_1,t_1)\ldots
g(T^n,\xi_n,t_n)] \rangle=0\quad\Rightarrow\quad [A(T,\xi,t),q_a]=0
\end{gather}
\reseteqn for the open string WZW correlators $A(T,\xi,t)$.

Continuing to follow Ref.~\cite{HO}, we next obtain the KZ-like
equations
\group{kz}
\begin{gather}
\begin{align}\nn
\pl_{t_i} A =& 2i\D_g(T^i)A -2i L_g^{ab}T_a^i A T^i_b\\ \nn
&+2iL_g^{ab} \sum_{j\ne i} \big{(} {A T_a^j T_b^i\over
1-e^{i(\phi_j-\phi_i)}} - {T_a^j A T_b^i\over
1-e^{i(\bar\phi_j-\phi_i)}}\\ &\quad\quad\quad\quad\quad\;- {T_a^i A
T_b^j \over 1-e^{i(\phi_j-\bar\phi_i)}} +{T_a^i T_b^j A \over
1-e^{i(\bar\phi_j-\bar\phi_i)}} \big{)}
\end{align}\\
\phi_i\equiv t_i+\xi_i \quad \bar\phi_i\equiv t_i-\xi_i,\quad
i,j=1,\ldots,n
\end{gather}
\reseteqn for the time dependence of the correlators $A\equiv
A(T,\xi,t)$ in open WZW theory. Tensor products are assumed in \eqref{kz}, 
as illustrated in the example:
\group{tensorproduct}
\begin{gather}
(A T^i T^j)_{\a_i\a_j}^{\;\;\;\b_i\b_j}\equiv (A T^i\otimes
T^j)_{\a_i\a_j}^{\;\;\;\b_i\b_j}=A_{\a_i\a_j}^{\;\;\;\c_i\c_j}(T^i)_{\c_i}{}^{\b_i}
(T^j)_{\c_j}{}^{\b_j}\\ [T_a^i,T_b^j]=i \d^{ij} f_{ab}{}^c T_c^i\ .
\end{gather}
\reseteqn To obtain these differential equations, we used the vertex
operator equation \eqref{voe1result}, the commutators
\group{Jpm-g}
\begin{gather}
[J_a^{(+)}(\pm\xi_i,t_i),g(T^j,\xi_j,t_j)]={g(T^j,\xi_j,t_j)T_a^j\over
1-e^{i(t_j+\xi_j-(t_i\pm\xi_i))}}- {T_a^j g(T^j,\xi_j,t_j)\over
1-e^{i(t_j-\xi_j-(t_i\pm\xi_i))}}\\
[J_a^{(-)}(\pm\xi_i,t_i),g(T^j,\xi_j,t_j)]={T_a^j
g(T^j,\xi_j,t_j)\over 1-e^{i(t_j-\xi_j-(t_i\pm\xi_i))}}-
{g(T^j,\xi_j,t_j)T_a^j\over 1-e^{i(t_j+\xi_j-(t_i\pm\xi_i))}}
\end{gather}
\reseteqn and the ground state condition \eqref{vacuum}.

The system \eqref{kz} resembles the usual \cite{KZ} KZ equations, but
it is in fact quite different:
\begin{itemize}
\item In \eqref{kz}, the variables $\phi_i$ and $\bar\phi_i$ are the
locations of the $i$-th non-abelian charge and the $i$-th generalized
non-abelian Dirichlet image charge respectively.
\item The ordinary KZ equations for the time dependence of the left and right mover
WZW correlators $A_+^\mathrm{WZW}$ and $A_-^\mathrm{WZW}$ on the cylinder are
\group{ordinarykz}
\begin{gather}
\begin{align}
\pl_{t_i} A_+^\mathrm{WZW} =& i\D_g(T^i)A_+^\mathrm{WZW} + A_+^\mathrm{WZW} \,2iL_g^{ab} \sum_{j\ne i} {T_a^j
T_b^i\over 1-e^{i(\phi_j-\phi_i)}}\\ \pl_{t_i} A_-^\mathrm{WZW} =& i\D_g(T^i)A_-^\mathrm{WZW} +
2iL_g^{ab} \sum_{j\ne i} {T_a^i T_b^j\over
1-e^{i(\bar\phi_j-\bar\phi_i)}}\, A_-^\mathrm{WZW} \ . 
\end{align}
\end{gather}
\reseteqn The coefficients of $A_\pm^\mathrm{WZW}$ in \eqref{ordinarykz} also appear in \eqref{kz}:
However in \eqref{kz} the $AT^j T^i$ terms represent the interactions among the non-abelian
charges, while the $T^i T^j A$ terms represent the interactions among the non-abelian
image charges.
\item In \eqref{kz}, the terms proportional to $T^j A T^i$ and 
$T^i A T^j$ are interactions between the charges and
the image charges. Such terms, with a simultaneous left and right
action, are unfamiliar in standard KZ theory.
\item The extra term $-2i L_g^{ab}T_a^i A T^i_b$ in \eqref{kz} will
be interpreted below.
\end{itemize}

It will be convenient to write the system \eqref{kz} in the more
conventional form 
\group{kz2}
\begin{gather}
\pl_{t_i} A = 2 i \D_g(T^i) A + A \o_i + \ob_i A - \o_i^a A T_a^i
-T_a^i A \ob_i^a \\ \o_i\equiv 2i L_g^{ab} \sum_{j\ne i} T_a^i T_b^j
f(\phi_j-\phi_i),\quad \ob_i\equiv 2i L_g^{ab} \sum_{j\ne i} T_a^i
T_b^j f(\bar\phi_j-\bar\phi_i)\label{firstconnections}\\ \o_i^a\equiv 2i L_g^{ab} \sum_{j}
T_b^j f(\bar\phi_j-\phi_i),\quad \ob_i^a\equiv 2i L_g^{ab} \sum_{j}
T_b^j f(\phi_j-\bar\phi_i)\label{allj}\\ f(x)\equiv{1\over
1-e^{ix}},\quad f(x)+f(-x)=1\label{fidentity}
\end{gather}
\reseteqn 
where we have used the identity \eqref{fidentity} to
re-express the $-2i L_g^{ab}T_a^i A T_b^i$ term in \eqref{kz}
as the $j=i$ terms of the completed sums in \eqref{allj}. In this
way, we interpret the $-2i L_g^{ab}T_a^i A T_b^i$ term in \eqref{kz}
as equivalent to two types of interaction between a given charge at
$\phi_i$ and its own image at $\bar\phi_i$. In what follows, the $\o$'s
of Eq.~\eqref{kz2} are referred to as connections.

We have checked
explicitly that these KZ-like differential equations satisfy the
appropriate integrability condition
\begin{equation}
[\pl_{t_i},\pl_{t_j}] A =0,\quad \forall \;i,j
\end{equation}
and that the differential equations are also consistent with the
$g$-global Ward identities \eqref{ward}. We will postpone the details
of this discussion, however, while we develop a simpler and more
comprehensive description of this system.

\vspace{-.07in}
\subsection{Constituent Vertex Operators}

In this subsection, we introduce constituent vertex operators
$g_\pm(T)$ which provide an enlightening alternative derivation of the
vertex operator differential equation \eqref{voe1}.

The constituent vertex operators are defined as
follows\footnote{Constituent vertex operators can also be introduced
in the same way for the classical theory, and all the same properties
obtained below for the quantum constituents can be obtained as well at
the classical level (see also Eq.~\eqref{chiralclassical}). In particular, we
will not need to know the explicit multiplication law for $g_-$ times
$g_+$, which presumably involves quantum groups.}:
\group{fact}
\begin{gather}
g(T,\xi,t) \equiv g_-(T,\xi,t) g_+(T,\xi,t)\\
\begin{align}
[J_a(m),g_+(T,\xi,t)]=&g_+(T,\xi,t)T_a e^{im(t+\xi)}\\
[J_a(m),g_-(T,\xi,t)]=&-T_a g_-(T,\xi,t) e^{im(t-\xi)}\ . 
\end{align}
\end{gather}
\reseteqn
Taken together, the simple commutation relations in \eqref{fact}
reproduce the algebra \eqref{modeJ-gq} of the current modes with the
full vertex operators $g(T)$.

By direct computation with \eqref{fact}, we find the simpler time
differential equations 
\group{factt}
\begin{gather}
\pl_t g_\pm(T,\xi,t)=i[H_g,g_\pm(T,\xi,t)]\\
\begin{align}
\pl_t g_+(T,\xi,t) =& 2i L^{ab}_g : J_a(\xi,t) g_+(T,\xi,t) T_b : + i
\D_g(T) g_+(T,\xi,t) \\ \pl_t g_-(T,\xi,t) =& - 2i L^{ab}_g :
J_a(-\xi,t) T_b g_-(T,\xi,t) : + i \D_g(T) g_-(T,\xi,t)
\end{align}
\end{gather}
\reseteqn 
for the constituent vertex operators. The normal ordering here is the same as in
\eqref{voe2} with $g\to g_\pm$.  Then we find that the time derivative
\eqref{voe1result} of the full $g(T)$ is a consequence of the constituent equations \eqref{factt} as follows:
\group{recompute}
\begin{gather}
\begin{align}
\pl_t g(T,\xi,t)=& \pl_t g_-(T,\xi,t) g_+(T,\xi,t) + g_-(T,\xi,t)
\pl_t g_+(T,\xi,t)\\\nn =&-2i L_g^{ab}\big{(} :\hspace{-.05in}
J_a(-\xi,t) T_b g_-(T,\xi,t)\hspace{-.05in}: g_+(T,\xi,t) -
g_-(T,\xi,t) :\hspace{-.05in} J_a(\xi,t) g_+(T,\xi,t) T_b
\hspace{-.05in}: \big{)}\\ &+2 i \D_g(T)
g(T,\xi,t)\label{nonordered}\\\nn =&2i L_g^{ab} : g(T,\xi,t)
J_a(\xi,t) T_b - J_a(-\xi,t) T_b g(T,\xi,t) :\\ &-2i L_g^{ab} T_a
g(T,\xi,t) T_b + 2 i \D_g(T) g(T,\xi,t)\ .\label{ordered}
\end{align}
\end{gather}
\reseteqn In this computation,
the expression in \eqref{nonordered} was not completely normal
ordered, so we used the equal-time commutators \group{complusminus}
\begin{gather}
[J_a^{(+)}(-\xi,t),g_+(T,\xi,t)] = {g_+(T,\xi,t) T_a \over 1-
e^{2i\xi}}\\ [J_a^{(-)}(\xi,t),g_-(T,\xi,t)] = {T_a g_-(T,\xi,t)\over
1- e^{-2i\xi}}\\ {1\over 1- e^{-2 i\xi}}+{1\over 1- e^{2 i \xi}}= 1
\end{gather}
\reseteqn to obtain the final completely normal ordered form in
\eqref{ordered}.

We emphasize that the $-2i L_g^{ab}T_a^i A T_b^i$ term in
Eq.~\eqref{voe1result} or \eqref{ordered} is a result of this final
normal ordering.

We also consider the action of the Virasoro generators $L_g(m)$ of the
affine-Sugawara construction \group{vir}
\begin{gather}
L_g(m)\equiv L_g^{ab}\sum_{p\in\mathbb{Z}}
:J_a(p)J_b(m-p):\label{virasoro1}\\
[L_g(m),L_g(n)]=(m-n)L_g(m+n)+{c_g\over12} m(m^2-1)\d_{m+n,0}
\end{gather}
\reseteqn on the constituent vertex operators.  By direct computation
with (\ref{fact}b-c) and (\ref{factt}b-c) we find that
\begin{equation}
[L_g(m),g_\pm(T,\xi,t)]= e^{im(t\pm\xi)} \big{(} -i\pl_t + m \D_g(T)
\big{)} g_\pm(T,\xi,t)\label{virasoro2}\ .
\end{equation}
We have checked that these commutators satisfy the $L,L,g_\pm$ Jacobi
identities.

\vspace{-.07in}
\subsection{The Constituents are Chiral}

We begin this discussion with the quantum version of the bulk momentum
operator \group{momentum}
\begin{gather}
\begin{align}
P_g(t)\equiv&\,\onetwopi \int_0^\pi d\xi \0b
L_g^{ab} :J_a(\xi,t)J_b(\xi,t)-J_a(-\xi,t)J_b(-\xi,t):\\
=&\,{i\over\pi}L_g^{ab}\sum_{m+n\ne 0} e^{-i(m+n)t} 
\left( {(-1)^{m+n}-1\over m+n} \right) :J_a(m)J_b(n):\\ =&-{2i\over\pi} \sum_{m\in\mathbb{Z}}
{e^{-i(2m+1)t}\over 2m+1} L_g(2m+1)
\end{align}\\
\pl_t P_g(t)={L_g^{ab}\over\pi}
:J_a(\pi,t)J_b(\pi,t)-J_a(0,t)J_b(0,t):\\
L_g(m)^\dagger=L_g(-m)\quad\Rightarrow \quad P_g(t)^\dagger=P_g(t)
\end{gather}
\reseteqn where $L_g(m)$ are the Virasoro generators in
\eqref{virasoro1}. By direct computation with \eqref{virasoro2}, we
find that \group{momentum2}
\begin{gather}
\begin{align} 
i[P_g(t),g_+(T,\xi,t)]=&4\big{(}\hspace{-.05in}\int_0^\xi d\eta \0b
e^{i\eta} \d(2\eta)\big{)} \pl_t g_+(T,\xi,t) + 4 e^{i\xi} \d(2\xi)
\D_g(T)g_+(T,\xi,t)\\
i[P_g(t),g_-(T,\xi,t)]=&-4\big{(}\hspace{-.05in}\int_0^\xi d\eta \0b
e^{-i\eta} \d(2\eta)\big{)} \pl_t g_-(T,\xi,t)+4
e^{-i\xi}\d(2\xi)\D_g(T)g_-(T,\xi,t)
\end{align}\\
4\int_0^\xi d\eta \0b e^{i\eta} \d(2\eta)=4\int_0^\xi d\eta \0b
e^{-i\eta} \d(2\eta)= \left\{\begin{array}{ll} 1 & \mbox{if
$0<\xi<\pi$} \\ 0 & \mbox{if $\xi=0,\pi.$} \end{array} \right .
\end{gather}
\reseteqn The summation identities
\begin{equation}
\sum_{n\in\mathbb{Z}} {e^{\pm i (2n+1)\xi}\over 2n+1} = \pm 2\pi i
\int_0^\xi d\eta \0b e^{\pm i\eta} \d(2\eta)
\end{equation}
were used to obtain these results.

The last terms in (\ref{momentum2}a-b) are quantum effects which
contribute only at the boundary, so that
\begin{equation}
\pl_\xi g_\pm(T,\xi,t) = i [P_g(t),g_\pm(T,\xi,t)]= \pm \pl_t
g_\pm(T,\xi,t)\quad 0<\xi<\pi
\end{equation}
is obtained in the bulk. Following our classical
intuition, we extend this result smoothly to include the boundary
\begin{equation}
\pl_- g_+(T,\xi,t) = \pl_+ g_-(T,\xi,t)=0,\quad
\pl_\pm=\pl_t\pm\pl_\xi,\quad 0\le\xi\le\pi
\label{chiral}
\end{equation}
which tells us that the constituent vertex operators $g_+(T)$ and
$g_-(T)$ are respectively chiral and antichiral.

It is simple to check that this extension to the boundary is
consistent: First, the chirality of $g_\pm(T)$ in the form
\begin{equation}
\half \pl_\pm g_\pm(T,\xi,t) = \pl_t g_\pm(T,\xi,t)
\label{chiralityform}
\end{equation}
allows us to rewrite the equations for $\pl_t g_\pm$ in
(\ref{factt}b-c) as light-cone differential equations for the
constituent vertex operators \group{partialpm}
\begin{gather}
\begin{align}
\half \pl_+ g_+(T,\xi,t) =& 2i L_g^{ab} : J_a(\xi,t) g_+(T,\xi,t) T_b
: + i \D_g(T) g_+(T,\xi,t) \\ \half \pl_- g_-(T,\xi,t) =& - 2i
L_g^{ab} : J_a(-\xi,t) T_b g_-(T,\xi,t) : + i \D_g(T) g_-(T,\xi,t) \ .
\end{align}
\end{gather}
\reseteqn Then one easily checks that these equations are consistent
with the chirality conditions \eqref{chiral}
\begin{equation}
\pl_+ \pl_- g_\pm(T,\xi,t)=\pl_- \pl_+ g_\pm(T,\xi,t)=0
\end{equation}
because $g_+(T)$, $J(+\xi)$ are chiral and $g_-(T)$, $J(-\xi)$ are
antichiral.

As another check on the consistency of the chiralities
\eqref{chiral}, we remark that the relations in \eqref{fact},
\eqref{chiral} and \eqref{partialpm} are nothing but the quantum
versions of the classical relations \group{chiralclassical}
\begin{gather}
g(T,\xi,t) \equiv g_-(T,\xi,t) g_+(T,\xi,t)\label{chiralclassical1}\\
\begin{align}
\{J_a(\pm\xi,t),g_+(T,\eta,t)\}=& 2\pi \d(\xi\mp\eta)g_+(T,\xi,t)T_a\\
\{J_a(\pm\xi,t),g_-(T,\xi,t)\}=&-2\pi \d(\xi\pm\eta)T_a g_-(T,\xi,t)
\end{align}\\
\pl_- g_+(T,\xi,t) = \pl_+ g_-(T,\xi,t)=0,\quad 0\le\xi\le\pi \\
\begin{align}
\pl_+ g_+(T,\xi,t) =& 2 i g_+(T,\xi,t) J(T,\xi,t)\\ \pl_- g_-(T,\xi,t)
=& 2 i J(T,-\xi,t) g_-(T,\xi,t)
\end{align}
\end{gather}
\reseteqn which should be taken to supplement our classical discussion above. As in
Ref.~\cite{Wi}, the relations \eqref{chiralclassical1} and
(\ref{chiralclassical}d-f) solve the classical relations for $\pl_\pm
g(T)$ in \eqref{classicalpm}.

The constituent vertex operator equations \eqref{partialpm} are
cylindrical analogues of the vertex operator equations for the left
and right mover vertex operators in ordinary WZW theory (see e.g. Ref.~\cite{HO}), and indeed
these equations guarantee that $g_+$ correlators and $g_-$
correlators
\begin{equation}
g_\pm(i)\equiv g_\pm(T^i,\xi_i,t_i),\quad A_\pm \equiv \langle
g_\pm(1)\ldots g_\pm(n) \rangle
\end{equation}
obey cylindrical analogues of the ordinary KZ equations
\group{kzanalogue}
\begin{gather}
\begin{align}
\half \pl_{i +} A_+ =& i\D_g(T^i) A_+ + A_+ \o_i\\ \half \pl_{i -} A_-
=& i\D_g(T^i) A_- + \ob_i A_-
\end{align}\\
\pl_{i -} A_+=\pl_{i +} A_-=0\\ \pl_{i +}\equiv \pl_{t_i} +
\pl_{\xi_i} \quad \pl_{i -}\equiv \pl_{t_i} - \pl_{\xi_i}\ .
\end{gather}
\reseteqn The connections $\o_i,\ob_i$ are defined in Eq.~\eqref{firstconnections}.
The commutator identities 
\group{commpm}
\begin{gather}
\begin{align}
[J_a^{(+)}(\pm\xi_i,t_i),g_+(T^j,\xi_j,t_j)]=&{g_+(T^j,\xi_j,t_j)T_a^j\over
1-e^{i(t_j+\xi_j-(t_i\pm\xi_i))}}\\
[J_a^{(+)}(\pm\xi_i,t_i),g_-(T^j,\xi_j,t_j)]=&-{T_a^j
g_-(T^j,\xi_j,t_j)\over 1-e^{i(t_j-\xi_j-(t_i\pm\xi_i))}}\\
[J_a^{(-)}(\pm\xi_i,t_i),g_+(T^j,\xi_j,t_j)]=&-{g_+(T^j,\xi_j,t_j)T_a^j\over
1-e^{i(t_j+\xi_j-(t_i\pm\xi_i))}}\\
[J_a^{(-)}(\pm\xi_i,t_i),g_-(T^j,\xi_j,t_j)]=&{T_a^j
g_-(T^j,\xi_j,t_j)\over 1-e^{i(t_j-\xi_j-(t_i\pm\xi_i))}}
\end{align}
\end{gather}
\reseteqn 
are needed in the derivation of \eqref{kzanalogue}.
Because the correlators $A_\pm$ are chiral, the equations in
\eqref{kzanalogue} are identical to the cylindrical KZ equations
given in \eqref{ordinarykz}.

But in open WZW theory, the full correlators $A$ of the full vertex
operators $g$ cannot be factorized into the constituent correlators
$A_\pm$: \group{nofact}
\begin{gather}
g(i)\equiv g(T^i,\xi_i,t_i)=g_-(i) g_+(i)\\ A=\langle g(1)\ldots
g(n)\rangle \ne \langle g_-(1)\ldots g_-(n)\rangle \langle
g_+(1)\ldots g_+(n)\rangle=A_- A_+ \ .
\end{gather}
\reseteqn This follows because, in open string theory, the current
modes $J(m)$ have non-trivial action on both $g_+$ and $g_-$ so that
$g_+$ and $g_-$ do not form independent subspaces \cite{HO} as in the
WZW model. Looking back, this phenomenon can also be understood as 
the fact that the strip current $J(\xi,t)$ does not commute  with the 
strip current $J(-\xi,t)$ at the boundary.

Finally, we may use Eq.~\eqref{chiralityform} to recast
\eqref{virasoro2} in the form
\begin{equation}
[L_g(m),g_\pm(T,\xi,t)]=e^{i(t\pm\xi)}\big{(}-{i\over2} \pl_\pm + m
\D_g(T) \big{)} g_\pm(T,\xi,t)
\label{nonchiralvir}
\end{equation}
and this form also satisfies the $L,L,,g_\pm$ Jacobi identities. Here
$\{L_g(m)\}$ acts on $g_+(T)$ as a left mover Virasoro acts on a left
mover Virasoro primary, but $\{L_g(m)\}$ also acts on $g_-(T)$ as a
right mover Virasoro $\{\bar{L}_g(m)\}$ acts on a right mover Virasoro
primary.

\vspace{-.07in}
\subsection{The Full Open String Vertex Operator Equations}

In this subsection, we use the results above for the chiral vertex
operators $g_\pm(T)$ to find the differential equations for the full vertex
operators $g(T)$.

Our first step is to use \eqref{chiral} and \eqref{partialpm} to
obtain the light-cone differential equations for $g(T)$
\group{partialg}
\begin{gather}
\begin{align}
\half \pl_+ g(T,\xi,t)=& 2i L_g^{ab} :J_a(\xi) g(T,\xi,t) T_b: - 2i
L_g^{ab}{T_a g(T,\xi,t) T_b\over 1-e^{-2 i \xi}} + i \D_g(T)
g(T,\xi,t) \\ \half \pl_- g(T,\xi,t)=& - 2i L_g^{ab} :J_a(-\xi) T_b
g(T,\xi,t): - 2i L_g^{ab}{T_a g(T,\xi,t) T_b\over 1-e^{2 i \xi}} + i
\D_g(T) g(T,\xi,t) \ .
\end{align}
\end{gather}
\reseteqn In the final step of this computation, the commutators
\eqref{complusminus} are needed again to obtain the fully normal
ordered form.  This result is the quantum version of the classical
result in Eq.~\eqref{classicalpm}: The normal ordered terms have the
same form as the classical result, and the remaining terms are quantum
effects from the normal ordering. The consistency of the system
\eqref{partialg}
\begin{equation}
[\pl_+,\pl_-] g(T,\xi,t)=0
\end{equation}
follows by construction from $g=g_- g_+$.

Taking linear combinations of Eqs.~(\ref{partialg}a-b), we also find
the $\pl_t$ and $\pl_\xi$ equations for the vertex operators $g(T)$:
\group{t-xi}
\begin{gather}
\begin{align}\nn
\pl_t g(T,\xi,t)=& 2i L_g^{ab} :J_a(\xi) g(T,\xi,t) T_b - J_a(-\xi)
T_b g(T,\xi,t) :\\ &- 2i L_g^{ab} T_a g(T,\xi,t) T_b + 2 i \D_g(T)
g(T,\xi,t) \label{agrees} \\ \nn \pl_\xi g(T,\xi,t)=& 2i L_g^{ab}
:J_a(\xi) g(T,\xi,t) T_b + J_a(-\xi) T_b g(T,\xi,t) :\\ &- 2 L_g^{ab}
T_a g(T,\xi,t) T_b \cot{\xi}
\end{align}\\
{1\over 1- e^{2i\xi}}+{1\over 1- e^{-2i\xi}}=1, \quad {1\over 1-
e^{2i\xi}} - {1\over 1- e^{-2i\xi}}=i\cot{\xi}\label{cot}
\end{gather}
\reseteqn where we have used the relations in \eqref{cot}. The
equation in \eqref{agrees} agrees of course with the earlier result
in \eqref{voe1}.

As a simple application, we use the vertex operator equations
\eqref{t-xi}, the ground state conditions \eqref{vacuum}
and the $g$-global Ward identity to compute the one-point 
correlators of the open WZW string 
\group{onetxieq}
\begin{gather}
[T_a,\langle g(T,\xi,t) \rangle ] = 0\\ \pl_t \langle g(T,\xi,t)
\rangle =0, \quad \pl_\xi \langle g(T,\xi,t) \rangle =
-2\D_g(T)\cot{(\xi)} \langle g(T,\xi,t) \rangle\ .
\end{gather}
\reseteqn
The solution of \eqref{onetxieq} is
\begin{equation}
 \langle g(T,\xi,t) \rangle = \one\, C(T) (\sin{\xi})^{-2\D_g(T)}
 \label{onetxi}
\end{equation}
where $C(T)$ is an undetermined number. As another application,
the vertex operator equations \eqref{partialg} or \eqref{t-xi} can be
used to study the operator product $g(1) g(2)$ and its operator
product expansion.

\vspace{-.07in}
\subsection{The Full Open String KZ Equations}

Using \eqref{vacuum}, \eqref{Jpm-g} and the vertex operator equations
\eqref{partialg}, we now obtain the full open string KZ equations
\group{kzpm}
\begin{gather}
A=\langle g(T^1,\xi_1,t_1)\ldots g(T^n,\xi_n,t_n) \rangle\\
\begin{align}
\pl_{\phi_i} A =& i\D_g(T^i) A + A \o_i - \o_i^a A T_a^i\\
\pl_{\bar{\phi}_i} A =& i\D_g(T^i) A + \ob_i A - T_a^i A \ob_i^a
\end{align}\\
\phi_i=t_i+\xi_i,\quad \bar{\phi}_i=t_i-\xi_i,\quad i=1,\ldots,n\\
\o_i\equiv 2i L_g^{ab} \sum_{j\ne i} T_a^i T_b^j
f(\phi_j-\phi_i),\quad \ob_i\equiv 2i L_g^{ab} \sum_{j\ne i} T_a^i
T_b^j f(\bar\phi_j-\bar\phi_i)\\ \o_i^a\equiv 2i L_g^{ab} \sum_{j}
T_b^j f(\bar\phi_j-\phi_i),\quad \ob_i^a\equiv 2i L_g^{ab} \sum_{j}
T_b^j f(\phi_j-\bar\phi_i)\\ f(x)\equiv{1\over 1-e^{ix}} \\
[\sum_{i=1}^n T_a^i,A]=0
\end{gather}
\reseteqn
for the correlators $A$ of open WZW theory. This system is a central result of this paper.

The $n$-point correlators in this system satisfy $2n$ partial differential equations 
in the $2n$ independent variables $\{\phi_i,\bar{\phi}_i\}$, 
so the complexity of the $n$-point correlators in open WZW string theory
is comparable to the complexity of the $2n$-point correlators of the ordinary
KZ equations. The solution for the open string one-point correlators is given in
\eqref{onetxi}. 

Because the form of this system is unfamiliar, we have checked its
integrability conditions carefully. Aside from considerable algebra
using the form of the connections $\o$, 
the only identities needed are \group{fid}
\begin{gather}
f(x-y)f(z-x)+f(z-y)f(y-x)-f(z-y)f(z-x)=0,\quad\forall\;x,y,z\\
f(x)+f(-x)=1\ .
\end{gather}
\reseteqn In detail, we find that the integrability conditions are
satisfied in the following way: \group{consistencyl-l}
\begin{gather}
[\pl_{\phi_i},\pl_{\phi_j}]A=0 \quad \mathrm{because:}\\
\pl_{\phi_i}\o_j - \pl_{\phi_j}\o_i=0,\quad \pl_{\phi_i}\o_j^a D
T_a^j-\pl_{\phi_j}\o_i^a D T_a^i=0\\ [\o_i,\o_j]=0,\quad
[\o_i^a,\o_j^b] D T_b^j T_a^i + \o_j^a D [\o_i,T_a^j] - \o_i^a D
[\o_j,T_a^i]=0,\quad \forall\; i,j
\end{gather}
\reseteqn

\group{consistencyr-r}
\begin{gather}
[\pl_{\bar{\phi}_i},\pl_{\bar{\phi}_j}]A=0 \quad \mathrm{because:}\\
\pl_{\bar{\phi}_i}\ob_j - \pl_{\bar{\phi}_j}\ob_i=0,\quad T_a^j D
\pl_{\bar{\phi}_i}\ob_j^a - T_a^i D \pl_{\bar{\phi}_j}\ob_i^a =0\\
[\ob_i,\ob_j]=0,\quad T_a^i T_b^j D [\ob_i^a,\ob_j^b] + [\ob_i,T_a^j]
D \ob_j^a - [\ob_j,T_a^i] D \ob_i^a=0,\quad\forall\; i,j
\end{gather}
\reseteqn

\group{consistencyl-r}
\begin{gather}
[\pl_{\phi_i},\pl_{\bar{\phi}_j}]A=0 \quad \mathrm{because:}\\
\pl_{\phi_i}\ob_j=\pl_{\bar{\phi}_j}\o_i=0,\quad T_a^j D \pl_{\phi_i}
\ob_j^a - \pl_{\bar{\phi}_j} \o_i^a D T_a^i=0\\ [\o_i^a,\ob_j] D T_a^i
- T_a^j D [\o_i,\ob_j^a] + T_a^j \o_i^b D T_b^i \ob_j^a - \o_i^b T_a^j
D \ob_j^a T_b^i =0,\quad \forall\;i,j \ .
\end{gather}
\reseteqn Here, $D$ is an arbitrary square matrix in the space of
correlators.

The compatibility between the open string KZ equations \eqref{kzpm}
and the $g$-global Ward identity \eqref{ward} 
\group{consistencyward}
\begin{gather}
q_a=\sum_i T_a^i,\quad [q_a,\o_i]=[q_a,\ob_i]=0\\ [q_a,T_b^i] D \o_i^b
+ T_b^i D [q_a,\o_i^b]=\ob_i^b D [q_a,T_b^i]+ [q_a,\ob_i^b] D T_b^i
=0,\quad\forall\;i
\end{gather}
\reseteqn can also be checked in the same way.

We also give the full open string KZ equations in the alternate form

\group{summarykz}
\begin{gather}
\begin{align}
\pl_{t_i} A =& 2 i \D_g(T^i) A + A \o_i + \ob_i A - \o_i^a A T_a^i
-T_a^i A \ob_i^a\\ \pl_{\xi_i} A =& A \o_i - \ob_i A - \o_i^a A T_a^i
+ T_a^i A \ob_i^a,\quad i=1,\ldots,n
\end{align}
\end{gather}
\reseteqn where the derivatives are now with respect to the basic
world-sheet variables $\{t_i,\xi_i\}$. In this system,
Eq.~(\ref{summarykz}a) is the same as \eqref{kz2}.

Still another form of our open string KZ system is given in App.~C,
where we use dual matrix representations to present 
the equations for the open $n$-point correlators as a single
``chiral'' KZ system in $2n$ variables.

Finally, we record the action of the Virasoro generators $L_g(m)$ on
the full vertex operators $g(T)$ \group{virg}
\begin{gather}
\begin{align}
[L_g(m), g(T,\xi,t)] =& \big{(} e^{im\bar{\phi}} (-i\pl_{\bar{\phi}} +
m \D_g(T)) + e^{im\phi}( -i\pl_{\phi} + m \D_g(T))\big{)} g(T,\xi,t)\\
=& e^{imt}\big{(} \cos(m\xi)(-i\pl_t +2m \D_g(T))+\sin(m\xi)\pl_\xi
\big{)} g(T,\xi,t)\ .
\end{align}
\label{L_g(m)-g}
\end{gather}
\reseteqn The result follows from Eq.~\eqref{nonchiralvir} and the chirality
\eqref{chiral} of the constituents.

\vspace{-.07in}
\subsection{General Open String CFT}

In this subsection, we extend our discussion to general open string
conformal field theory, using what we now know about open WZW theory as a
model. In particular, we assume that the general open string CFT is
governed by a single set of Virasoro generators $L(m)$ \group{generalvir}
\begin{gather}
[L(m),L(n)]=(m-n)L(m+n)+{c\over 12} m(m^2-1) \d_{m+n,0}\\ L(|m|\le 1)
|0 \rangle = \langle 0 | L(|m|\le 1) =0
\end{gather}
\reseteqn with an $SL(2)$ invariant ground state $|0\rangle$. For
open CFT's based on a current algebra, $|0\rangle$ is the
affine ground state. The set of all open CFT's is very large, including open 
analogues of coset constructions \cite{BH,Ha,GKO,Rev}, affine-Virasoro 
constructions \cite{HK,MP,Rev} and conformal sigma models. 

In each open CFT, we also consider the set of open string 
Virasoro primary fields $\{\Phi_i\}$, which are defined to
satisfy
\begin{equation}
[L(m), \Phi_i(\xi,t)] = e^{imt}\big{(} \cos(m\xi)(-i\pl_t +2m
\D_i)+\sin(m\xi)\pl_\xi \big{)} \Phi_i(\xi,t) \ .
\label{virphi}
\end{equation}
Open string Virasoro quasiprimary fields  $\chi_i(\xi,t)$ are defined to 
satisfy \eqref{virphi} for $|m|\le 1$. 

Then the $SL(2)$ Ward identities 
\group{sl2ward}
\begin{gather}
A\equiv \langle \chi_1 \ldots \chi_n \rangle\\
\begin{align}
L(0):& \quad \sum_{i=1}^n \pl_{t_i} A =0\label{timeinv}\\ L(\pm 1):&
\quad \sum_{i=1}^n e^{\pm i t_i} \big{(} \cos{\xi_i} (-i \pl_{t_i} \pm
2 \D_i) \pm \sin{\xi_i}\pl_{\xi_i}) \big{)} A =0
\end{align}
\end{gather}
\reseteqn follow in the usual way for open string correlators
of sets of Virasoro quasiprimaries. The relation \eqref{timeinv} 
expresses the time translation invariance of the correlators.

The open string WZW vertex operators $g(T)$ above are examples of open string Virasoro primary
fields with $\D_i=\D_g(T^i)$ and, indeed, the $SL(2)$ Ward identities 
\eqref{sl2ward} are satisfied by the correlators of the open
string KZ system \eqref{kzpm}. In particular,
we have checked that the compatibility conditions
\group{sl2consistency}
\begin{gather}
\begin{align}
L_g(0)\!:& \; \sum_{i=1}^n (2 i \D_g(T^i) A + A \o_i + \ob_i A -
\o_i^a A T_a^i -T_a^i A \ob_i^a )\! = \!0\\ L_g(-1)\!:& \;
\sum_{i=1}^n \big{(} e^{- i(t_i-\xi_i)}(\ob_i A-T_a^i A \ob_i^a) +
e^{- i(t_i+\xi_i)}(A \o_i- \o_i^a A T_a^i) \big{)} \! = \! 0\\
L_g(+1)\!:& \; \sum_{i=1}^n \big{(} e^{i(t_i-\xi_i)}(i\D_g(T^i)A+\ob_i
A-T_a^i A \ob_i^a) + e^{i(t_i+\xi_i)}(i\D_g(T^i)A +A \o_i- \o_i^a A
T_a^i) \big{)}\! = \! 0
\end{align}
\end{gather}
\reseteqn are satisfied by the connections $\o$, using also the $g$-global 
Ward identity for (\ref{sl2consistency}a) and (\ref{sl2consistency}c).

The $SL(2)$ Ward identities \eqref{sl2ward} can be put in a more recognizable form
\group{standardsl2}
\begin{gather}
A\equiv \left(\prod_{i=1}^n (z_i \bar{z}_i)^{\D_i}\right) F\\
\begin{align}
L(-1):&\quad \sum_{i=1}^n (\pl_i+\bar{\pl}_i)F=0\\
L(0):&\quad \sum_{i=1}^n ((z_i\pl_i+\D_i)+(\bar{z}_i\bar{\pl}_i+\D_i)) F=0\\
L(1):&\quad \sum_{i=1}^n (z_i(z_i\pl_i+2\D_i)+\bar{z}_i(\bar{z}_i\bar{\pl}_i+2\D_i)) F=0
\end{align}\\
z_i\equiv e^{i(t_i+\xi_i)},\quad \bar{z}_i\equiv e^{i(t_i-\xi_i)},\quad 
\pl_i\equiv {\pl\over\pl z_i},\quad \bar{\pl}_i\equiv {\pl\over\pl \bar{z}_i}
\end{gather}
\reseteqn
so the $SL(2)$ Ward identities for the open string $n$-point $F$ factor have the form of 
the ordinary $SL(2)$ Ward identities, now for a correlator with $2n$ 
points $z_1,\bar{z}_1\ldots z_n\bar{z}_n$. The solutions to these 
equations are easily read off from the general solution of $SL(2)$ Ward 
identities given in Ref.~\cite{BPZ}. 

As a simple example, we find for the open string one-point correlators
\begin{equation}
\langle \chi_i(\xi_i,t_i) \rangle \propto{(z_i \bar{z}_i)^{\D_i}\over (z_i-\bar{z}_i)^{2\D_i}}\propto 
{1\over (\sin{\xi_i})^{2\D_i}}
\label{onesl2}
\end{equation}
in agreement with our solution \eqref{onetxi} of the open string KZ
equations and the $g$-global Ward identity. In the solution 
\eqref{onesl2}, the open string one-point $F$ factor 
\begin{equation}
F \propto {1\over (z_i - \bar{z}_i)^{2\D_i}}
\end{equation}
has the form of the usual two-point correlator between a closed string 
quasiprimary field at $z_i$ and another closed string quasiprimary field 
at a point called $\bar{z}_i$: In open string theory, however, the points 
$z_i$ and $\bar{z}_i$ are the locations of the charge and the image charge respectively.
 
The $SL(2)$ forms of the open string correlators will be helpful 
in solving the open string KZ equations for the
multi-point correlators on the strip.

\bigskip
\noindent
{\bf Acknowledgements} 

For helpful discussions, we thank N. Arkani-Hamed, P. Horava,
A. Konechny, N. Resheti\-kin, C. Schwei\-gert and K. Skenderis.

After completion of this work, a paper ``Canonical Quantization of the
Boundary Wess-Zumino-Witten Model'', hep-th/0101170 appeared on the net 
by Gawedzki, Todorov and Tran-Ngoc-Bich. This paper also considers
the dynamics of open WZW strings, and, although the approach is quite different,
there is an explicit overlap with our result
Eq.~\eqref{J-gq}.

This work was supported in part by the Director, Office of Energy
Research, Office of High Energy and Nuclear Physics, Division of High
Energy Physics of the U.S. Department of Energy under Contract
DE-AC03-76SF00098 and in part by the National Science Foundation under
grant PHY95-14797. The work of S.G. was partially supported by the
European commission RTN programme HPRN-CT-2000-00131.

\appendices
\vspace{-.05in}


\app{abelian}{The Abelian Case}
In order to compare with the discussion in the text, we consider the
classical action formulation of the simplest abelian Dirichlet string:
\namegroup{abelianaction} 
\alpheqn
\begin{gather}
S=\int_0^\pi d\xi \0b \mathcal{L}\\ \mathcal{L} = {1\over 8\pi} \pl_+
x\, \pl_- x = {1\over 8 \pi} ((\pl_t x)^2 - (\pl_\xi x)^2)\\
(\pl_t^2-\pl_\xi^2)x(\xi,t)=0,\quad \pl_t x(0,t)=\pl_t x(\pi,t)=0\ .
\end{gather}
\reseteqn The solution of (\ref{abelianaction}c) is
\namegroup{solution} 
\alpheqn
\begin{gather}
\half x(\xi,t)=q + J(0) \xi + \sum_{m\ne 0} J(m) {\sin(m\xi)\over m}
e^{-imt}\label{abeliancoordinate}\\
\begin{align}
\pl_t x(\xi,t)=& -2 i \sum_{m\in\mathbb{Z}} J(m) \sin(m\xi)
e^{-imt}=J(\xi,t)-J(-\xi,t)\\ \pl_\xi x(\xi,t)=& 2
\sum_{m\in\mathbb{Z}} J(m) \cos(m\xi) e^{-imt}=J(\xi,t)+J(-\xi,t)
\end{align}
\end{gather}
\reseteqn where the current modes $J(m)$ satisfy the
usual abelian current algebra 
\namegroup{abeliancurrent} 
\alpheqn
\begin{gather}
J(\xi,t)=\sum_{m\in\mathbb{Z}} J(m)e^{-im(t+\xi)},\quad
J(-\xi,t)=\sum_{m\in\mathbb{Z}} J(m)e^{-im(t-\xi)}\\
\begin{align}
\{ J(\xi,t), J(\eta,t)\} =& 2\pi i \pl_\xi\d(\xi-\eta) \\ \{J(\xi,t),
J(-\eta,t)\} =& 2\pi i \pl_\xi\d(\xi+\eta) \\ \{J(-\xi,t),
J(-\eta,t)\} =& - 2\pi i \pl_\xi\d(\xi-\eta)
\end{align} \\
\{J(m),J(n)\}=m\d_{m+n,0}\ .
\end{gather}
\reseteqn 
We also assume that 
\begin{equation}
\{q,J(m\ne 0)\}=0
\end{equation}
but we leave the bracket $\{q,J(0)\}$ undetermined for the moment.
The string coordinate $x(\xi,t)$ can also be
written as
\begin{equation}
x(\xi,t)=2q + \big{(} J(0) (t+\xi) + i \sum_{m\ne 0} J(m)
{e^{-im(t+\xi)}\over m} \big{)} - \big{(} J(0) (t-\xi) + i \sum_{m\ne
0} J(m) {e^{-im(t-\xi)}\over m} \big{)}
\end{equation}
and exponentials of $x(\xi,t)$ show the abelian analogue of the right
and left mover product structure of the non-abelian vertex operator 
$g=g_-g_+$ discussed in the text.

The momentum $p$ canonical to $x$ follows from the Lagrange density
\eqref{abelianaction}:
\begin{equation}
p(\xi,t)\equiv \onefourpi \pl_t x(\xi,t)=\onefourpi
(J(\xi,t)-J(-\xi,t))\ .
\label{abelianmomentum}
\end{equation}
Using the forms (\ref{solution}c) and 
\eqref{abelianmomentum} for $\pl_\xi x$ and $p$, we find the phase space realization of the
currents 
\namegroup{abelianbowcock} 
\alpheqn
\begin{gather}
J(\xi,t)= 2\p p(\xi)+\half \pl_\xi x(\xi),\quad J(-\xi,t)= -2\p
p(\xi)+\half \pl_\xi x(\xi)
\end{gather}
\reseteqn 
and from the current algebra 
\eqref{abeliancurrent}, we compute the phase space brackets
\namegroup{abeliancommutators} \alpheqn
\begin{gather}
\begin{align}
\{x(\xi,t),x(\eta,t)\}=&4(\eta-\xi)\{q,J(0)\}\\
\{x(\xi,t),p(\eta,t)\}=&i(\d(\xi-\eta)-\d(\xi+\eta))\\
\{p(\xi,t),p(\eta,t)\}=&0
\end{align}\\
\begin{align}
\{J(\xi,t),x(\eta,t)\}=&-2\pi i
(\d(\xi-\eta)-\d(\xi+\eta))+2\{J(0),q\}\\ \{J(-\xi,t),x(\eta,t)\}=&
2\pi i (\d(\xi-\eta)-\d(\xi+\eta))+2\{J(0),q\}
\end{align}
\end{gather}
\reseteqn 
in terms of the unknown quantity $\{q,J(0)\}$. We remark in particular that
the second term in (\ref{abeliancommutators}b) corresponds to the 
presence of a Dirichlet image charge.

To determine the unknown quantity $\{q,J(0)\}$, we require the 
consistency of the 
action and Hamiltonian formulations of the system. The Hamiltonian of the Dirichlet string
follows by Legendre
transformation 
\namegroup{abelianham} \alpheqn
\begin{gather}
H = \int_{0}^{\pi} d\xi \0b \mathcal{H}\\ \mathcal{H}=\pl_t x P
-\mathcal{L}= 2\pi p^2+{1\over 8\pi} (\pl_\xi x)^2 =
\onefourpi(J^2(\xi)+J^2(-\xi)) 
\end{gather}
\reseteqn  and then we may recompute $\pl_t x$ from the Hamiltonian equations 
of motion
\begin{equation}
\pl_t x(\xi,t)=i\{H,x(\xi,t)\}=4\pi p(\xi,t)+ 2i \{J(0),q\} J(0)
\label{abelianxdot}
\end{equation}
using the brackets in \eqref{abeliancommutators}. For agreement with \eqref{abelianmomentum} we must
set
\begin{equation}
\{J(0),q\}=0
\label{assumption}
\end{equation}
and then all the previous relations of this appendix are in agreement with the abelian
limit \namegroup{abelianlimit} \alpheqn
\begin{gather}
x^i\to x,\quad p^i\to p,\quad B_{ij}\to 0\\ G_{ab}\to 1,\quad
f_{ab}{}^c \to 0,\quad e_i{}^a\to 1,\quad \eb_i{}^a\to -1
\end{gather}
\reseteqn of our discussion in the text.

In the abelian case, there is a second possible phase space
realization of the currents 
\begin{equation}
J(\xi,t)\equiv 2\p p(\xi)+\half \pl_\xi x(\xi),\quad J(-\xi,t)\equiv
2\p p(\xi)-\half \pl_\xi x(\xi)
\label{neumannbowcock}
\end{equation}
which differs from Eq.~\eqref{abelianbowcock} only by the
overall sign in the realization of $J(-\xi,t)$. Periodicity of the
cylinder current in this case gives the Neumann boundary conditions
\namegroup{neumannbc} \alpheqn
\begin{gather}
J(0,t)=J(-0,t),\quad J(\pi,t)=J(-\pi,t)\quad\mathrm{or}\\ \pl_\xi
x(0,t)=\pl_\xi x(\pi,t)=0\ .
\end{gather}
\reseteqn
The realization \eqref{neumannbowcock} must be taken with the same Hamiltonian
\eqref{abelianham} and the same current algebra
\eqref{abeliancurrent}. (The Neumann system is T-dual to the Dirichlet
system.) The solution to the equations of motion with Neumann boundary
conditions is
\begin{equation}
\half x(\xi,t)=q + J(0) t + i \sum_{m\ne 0} J(m) {\cos (m\xi)\over m}
e^{-imt}\ .
\label{neumannsolution}
\end{equation}
In this case all is consistent when
\begin{equation}
\{q,J(0)\}=i
\end{equation}
and then \eqref{neumannsolution} and \eqref{abeliancurrent} give
the phase space brackets 
\namegroup{neumanncommutators} \alpheqn
\begin{gather}
\begin{align}
\{x(\xi,t),x(\eta,t)\}=&0\\
\{x(\xi,t),p(\eta,t)\}=&i(\d(\xi-\eta)+\d(\xi+\eta))\\
\{p(\xi,t),p(\eta,t)\}=&0
\end{align}
\end{gather}
\reseteqn 
which show a Neumann image charge in the second term of
(\ref{neumanncommutators}b).

\vspace{-.05in}
\app{brackets}{The Remaining Phase Space Brackets}

In this appendix, we use the results of the text and the chain rule to
compute the rest of the phase space brackets of open WZW theory. In
particular we need Eqs.~\eqref{inverse}, \eqref{J-x} and
\eqref{xx-eqmatrix}, and the results are given in terms of the
coordinate brackets $\D=\{x,x\}$ in \eqref{xx-final}:
  
\begin{align}
\nonumber \{x^i(\xi,t),p_j(B,\eta,t)\} =& \{x^i(\xi),\onefourpi
\big{(} e(\eta)_j{}^a J_a(\eta) + \eb(\eta)_j{}^a J_a(-\eta)\big{)} \}
\\ \nonumber =& i \d_j{}^i \d(\xi-\eta) + {i\over 2} \big{(}
\eb(\eta)_j{}^a e(\xi)_a{}^i + e(\eta)_j{}^a \eb(\xi)_a{}^i \big{)}
\d(\xi+\eta) \\ &+ \onefourpi \D^{ik}(\xi,\eta) \big{(} \pl_k
e(\eta)_j{}^a J_a(\eta) + \pl_k \eb(\eta)_j{}^a J_a(-\eta) \big{)}
\label{x-pb}
\end{align}

\begin{align}
\nonumber \{x^i(\xi,t),p_j(\eta,t)\} =& \{x^i(\xi),p_j(B,\eta) -
\onefourpi B_{jk}(\eta) \pl_\eta x^k\} \\ \nonumber =& i \d_j{}^i
\d(\xi-\eta) + {i\over 2} \big{(} \eb(\eta)_j{}^a e(\xi)_a{}^i +
e(\eta)_j{}^a \eb(\xi)_a{}^i +\Psi^{ik}(\xi,\eta) B_{jk} (\eta)
\big{)} \d(\xi+\eta) \\ \nonumber &+ \onefourpi
\D^{ik}(\xi,\eta)\big{(}\pl_k e(\eta)_j{}^a J_a(\eta) + \pl_k
\eb(\eta)_j{}^a J_a(-\eta)\big{)}\\ &+ \onefourpi
\D^{ik}(\xi,\eta)\big{(} \L(\eta)_k{}^l B_{jl}(\eta) + \pl_k
B_{jl}(\eta) \pl_\eta x^l \big{)}
\label{x-p}
\end{align}

\begin{align}
\nonumber \{J_a(\xi,t), p_i(B,\eta,t) \} =& \{J_a(\xi), \onefourpi
\big{(} e(\eta)_i{}^b J_b(\eta) + \eb(\eta)_i{}^b J_b(-\eta)\big{)} \}
\\ \nonumber =& {i\over 2} e(\eta)_i{}^b G_{ba} \pl_\xi \d(\xi-\eta) +
2\pi i \pl_i e(\eta)_a{}^j p_j(B,\eta) \d(\xi-\eta) \\ &+ {i\over 2}
\eb(\eta)_i{}^b G_{ba} \pl_\xi \d(\xi+\eta) + 2\pi i \pl_i
\eb(\eta)_a{}^j p_j(B,\eta) \d(\xi+\eta)
\label{J-p}
\end{align}

\namegroup{pb-pb} \alpheqn
\begin{gather}
\begin{align}
\nonumber \{ p_i(B,\xi,t),p_j(B,\eta,t) \} =& \{\onefourpi \big{(}
e(\xi)_i{}^a J_a(\xi) + \eb(\xi)_i{}^a J_a(-\xi)\big{)},p_j(B,\eta)
\}\\ \nonumber =& -{i\over 4\pi} f_{ab}{}^d G_{dc} e_i{}^a e_j{}^b
e_k{}^c \pl_\xi x^k \d(\xi-\eta) \\ \nonumber &+ {i\over 8 \pi}
\big{(} e(\xi)_i{}^a G_{ab} \eb(\eta)_j{}^b - \eb(\xi)_i{}^a G_{ab}
e(\eta)_j{}^b \big{)}\pl_\xi\d(\xi+\eta) \\ \nonumber &+ {i\over 2}
\big{[} \big{(} e(\xi)_i{}^a \pl_j \eb(\eta)_a{}^k + \eb(\xi)_i{}^a
\pl_j e(\eta)_a{}^k \big{)} p_k(B,\eta) \\ \nonumber & - \big{(}
e(\eta)_j{}^a \pl_i \eb(\xi)_a{}^k + \eb(\eta)_j{}^a \pl_i
e(\xi)_a{}^k \big{)} p_k(B,\xi) \big{]} \d(\xi+\eta) \\ \nonumber &-
{i\over 8\pi} f_{ab}{}^c \big{(} J_c(\xi) e(\xi)_i{}^a \eb(\eta)_j{}^b
+ J_c(\eta) \eb(\xi)_i{}^a e(\eta)_j{}^b \big{)} \d(\xi+\eta) \\ &+
{1\over 16 \pi^2}\D^{kl}(\xi,\eta) \tilde{\L}_{ki}(\xi)
\tilde{\L}_{lj}(\eta) \ ;
\end{align} \\
\tilde{\L}_{ij}(\xi)\equiv \pl_i e(\xi)_j{}^a J_a(\xi) + \pl_i
\eb(\xi)_j{}^a J_a(-\xi) \ .
\end{gather}
\reseteqn The momentum bracket $\{p_i(\xi,t),p_j(\eta,t)\}$ follows
from Eq.~(\eqref{pb-pb}) and the definition (\eqref{pb}):
\begin{align}\nn
\{ p_i(\xi,t),p_j(\eta,t) \} =& {i\over 8 \pi} \big{(} e(\xi)_i{}^a
G_{ab} \eb(\eta)_j{}^b - \eb(\xi)_i{}^a G_{ab} e(\eta)_j{}^b \big{)}
\pl_\xi\d(\xi+\eta) \\ \nn &-{i\over 8\pi}\big{[} B_{ik}(\xi)\big{(}
\eb(\eta)_j{}^a e(\xi)_a{}^k + e(\eta)_j{}^a
\eb(\xi)_a{}^k\big{)}\\\nn &+\big{(} \eb(\eta)_i{}^a e(\xi)_a{}^k +
e(\eta)_i{}^a \eb(\xi)_a{}^k\big{)} B_{kj}(\eta) \big{]}
\pl_\xi\d(\xi+\eta)\\\nn &+ {i\over 2} \big{[} \big{(} e(\xi)_i{}^a
\pl_j \eb(\eta)_a{}^k + \eb(\xi)_i{}^a \pl_j e(\eta)_a{}^k \big{)}
p_k(B,\eta) \\ \nonumber & - \big{(} \pl_i \eb(\xi)_a{}^k
e(\eta)_j{}^a + \pl_i e(\xi)_a{}^k \eb(\eta)_j{}^a \big{)} p_k(B,\xi)
\big{]} \d(\xi+\eta) \\ \nonumber &- {i\over 8\pi} f_{ab}{}^c \big{(}
J_c(\xi) e(\xi)_i{}^a \eb(\eta)_j{}^b + J_c(\eta) \eb(\xi)_i{}^a
e(\eta)_j{}^b \big{)} \d(\xi+\eta) \\ \nn &+ {i\over 8\pi}
\big{[}\big{(} \eb(\xi)_i{}^a e(\eta)_a{}^l + e(\xi)_i{}^a
\eb(\eta)_a{}^l \big{)} \pl_l B_{jk}(\eta)\pl_\eta x^k(\eta)\\\nn
&-\pl_l B_{ik}(\xi)\pl_\xi x^k(\xi) \big{(} \eb(\eta)_j{}^a
e(\xi)_a{}^l + e(\eta)_j{}^a \eb(\xi)_a{}^l
\big{)}\big{]}\d(\xi+\eta)\\\nn &+ {i\over 8\pi}\big{[} \big{(}
\eb(\xi)_i{}^a \pl_k e(\eta)_a{}^l + e(\xi)_i{}^a \pl_k
\eb(\eta)_a{}^l \big{)} \pl_\eta x^k(\eta) B_{jl}(\eta)\\ \nn
&-B_{il}(\xi) \big{(} \eb(\eta)_j{}^a \pl_k e(\xi)_a{}^l +
e(\eta)_j{}^a \pl_k \eb(\xi)_a{}^l \big{)} \pl_\xi
x^k(\xi)\big{]}\d(\xi+\eta)\\ \nn &+ {1\over 16
\pi^2}\D^{kl}(\xi,\eta) \tilde{\L}_{ki}(\xi) \tilde{\L}_{lj}(\eta) \\
\nn &-\onefourpi B_{ik}(\xi) \pl_l B_{jm}(\eta) \pl_\eta
x^m(\eta)\big{(}2\pi i \Psi^{kl} (\xi,\eta) \d(\xi+\eta) +
\D^{kn}(\xi,\eta) \L(\eta)_n{}^l \big{)}\\ \nn &-\onefourpi \pl_k
B_{im}(\xi) \pl_\xi x^m(\xi) B_{jl}(\eta)\big{(}2\pi i \Psi^{kl}
(\xi,\eta) \d(\xi+\eta) + \L(\xi)_n{}^l \D^{kn}(\xi,\eta) \big{)}\\
\nn &+ {1\over 16 \pi^2}\pl_m B_{ik}(\xi) \pl_\xi x^k(\xi) \pl_n
B_{jl}(\eta) \pl_\eta x^l(\eta) \D^{mn}(\xi,\eta)\\ &+ {1\over 16
\pi^2}B_{ik}(\xi) B_{jl}(\eta) \pl_\xi \pl_\eta \D^{kl}(\xi,\eta) .
\label{p-p}
\end{align}
This final result for $\{p,p\}$ contains no bulk terms
(proportional to $\d(\xi-\eta)$), so it vanishes except at the
boundary.  Moreover, as in the case of $\{x,x\}$, this
non-commutativity is essentially non-abelian and vanishes 
(with $\D$) in the abelian limit.

\vspace{-.05in}
\app{kolia}{Presentation as a Single ``Chiral'' KZ system}

In this appendix, we present our open string KZ system for the open
$n$-point correlators $A$ as a single ``chiral'' KZ system in $2n$
variables $\mu$: \namegroup{2nvariables} \alpheqn
\begin{gather}
\mu \equiv \{\phi_i,\bar\phi_i\},\quad \pl_\mu\equiv\{ \pl_{\phi_i},
\pl_{\bar\phi_i}\}\\ \phi_i=t_i+\xi_i,\quad \bar\phi_i=t_i-\xi_i,\quad
\pl_{\phi_i}=\half \pl_{i +},\quad \pl_{\bar\phi_i}=\half \pl_{i-}\
.
\end{gather}
\reseteqn The variables $\phi_i$ and $\bar\phi_i$ were defined in
\eqref{kz}. The result is 
\namegroup{onechiral} 
\alpheqn
\begin{gather}
\pl_\mu \tilde{A}= \tilde{A} W_\mu\\
\begin{align}
W_{\phi_i}=& i\D_g(T^i) + 2 i L_g^{ab}\big{(} \sum_{j\ne i} T_a^i
T_b^j f(\phi_j-\phi_i) + \sum_{j} T_a^i \bar{T}_b^j
f(\bar\phi_j-\phi_i)\big{)}\\ W_{\bar\phi_i}=& i\D_g(T^i) + 2 i
L_g^{ab}\big{(} \sum_{j\ne i} \bar{T}_a^j \bar{T}_b^i
f(\bar\phi_j-\bar\phi_i) + \sum_{j} T_a^j \bar{T}_b^i
f(\phi_j-\bar\phi_i)\big{)}
\end{align}\\
\tilde{A} Q_a=0,\quad Q_a\equiv \sum_{i=1}^n (T_a^i+\bar{T}_a^i)
\end{gather}
\reseteqn where the function $f$ is defined in \eqref{kz2}. In
\eqref{onechiral} the matrices $T^i$ are associated to the charges at
$\phi_i$, and the matrices $\bar{T}^i$ are associated to the image
charges at $\bar\phi_i$. The matrix $\bar{T}^i$ is the
representation dual to the irrep $T^i$
\begin{equation}
\bar{T}_\a{}^\b\equiv - T_\b{}^\a\ .
\end{equation}
In this notation, the matrix $A$ of the text is treated as a single
large row $\tilde{A}$ and the matrices on the right of $\tilde{A}$ act 
to the left as a tensor product. As an example, we work out a 
representative term of the $\tilde{A}T^i\bar{T}^j$ type in \eqref{onechiral}, 
starting with the notation of the text, 
\namegroup{indices} \alpheqn
\begin{gather}
\begin{align} \nn
(T^i A T^j)_\a{}^\b = & (T^i)_\a{}^\c
A_\c{}^\d (T^j)_\d{}^\b\equiv - \tilde{A}^{\d\c} (T^j)_\d{}^\b
(\bar{T}^i)_\c{}^\a \\ \equiv& -(\tilde{A} T^j\otimes
\bar{T}^i)^{\b\a}\equiv-(\tilde{A} T^j \bar{T}^i)^{\b\a}
\end{align}\\
\tilde{A}^{\b\a}\equiv A_\a{}^\b
\end{gather}
\reseteqn where $A_\a{}^\b$ is the correlator \eqref{correlator} of the text. 
In this example, we have suppressed inactive indices for simplicity. More 
generally, one finds that 
\begin{equation}
\tilde{A}^{\b_1\ldots\b_n,\a_1\ldots\a_n}= A_{\a_1\ldots\a_n}^{\;\;\;\b_1\ldots\b_n}
\end{equation}
where, in the chiral form $\tilde{A}$, the $T$'s operate to the left on the $\b$ 
indices and the $\bar T$'s operate to the left on the $\a$ indices.

In 
this presentation one finds that the consistency conditions 
(\ref{consistencyl-l}-\ref{consistencyward}) take the simple form
\namegroup{finalcons}
\alpheqn
\begin{gather}
\pl_\mu W_\nu -\pl_\nu W_\mu = [W_\mu,W_\nu]=0\\
[Q_a,W_\mu]=0
\end{gather}
\reseteqn
so that the connection $W_\mu$ is abelian flat. 

The system \eqref{onechiral} looks even more familiar
\namegroup{reduced}
\alpheqn
\begin{gather}
\tilde{A}\equiv \prod_{i=1}^n (z_i \bar{z}_i)^{\D(T^i)} \tilde{F}\\
\pl_i \tilde{F} = \tilde{F}\,2 L_g^{ab}\big{(} \sum_{j\ne i} {T_a^i
T_b^j \over z_i - z_j} + \sum_{j} {T_a^i \bar{T}_b^j\over
z_i-\bar{z}_j} \big{)}\\
\bar{\pl}_i \tilde{F} = \tilde{F}\, 2 L_g^{ab}\big{(} \sum_{j\ne i} {\bar{T}_a^j
\bar{T}_b^i \over \bar{z}_i - \bar{z}_j} + \sum_{j} {T_a^j \bar{T}_b^i\over
\bar{z}_i-z_j} \big{)}\\
z_i=e^{i(t_i+\xi_i)},\quad \bar{z}_i=e^{i(t_i-\xi_i)},\quad \pl_i={\pl\over\pl z_i},\quad \bar{\pl}_i={\pl\over\pl \bar{z}_i}
\end{gather}
\reseteqn
in terms of the reduced correlators $\tilde{F}$ and the $z,\bar z$ variables
of Sec.~7.7.

This ``chiral'' description of our system is now in standard KZ form,
and one may apply standard \cite{Va} formal methods in KZ theory to
obtain solutions of our open string KZ system as integral representations. This appendix was
worked out in a conversation with N. Reshetikin.

\end{document}